\pdfoutput=1
\documentclass[sigplan,nonacm,10pt]{acmart}
\renewcommand\footnotetextcopyrightpermission[1]{}

\usepackage{xurl}
\usepackage{tikz}
\usetikzlibrary{positioning, fit, arrows.meta, backgrounds}
\usepackage{fvextra}
\usepackage{newfloat}
\DeclareFloatingEnvironment[fileext=lol,listname={List of Listings},name=Listing,placement=tb]{listing}

\usepackage{graphicx}
\usepackage{xspace}
\usepackage{listings}
\usepackage{xcolor}
\usepackage{multirow}
\usepackage{subcaption}
\usepackage{paralist}
\usepackage{enumitem}
\setlist{
  topsep=6pt,
  itemsep=1pt,
  parsep=0pt,
  partopsep=0pt
}

\newcommand{\eg}{e.g.,\xspace}

\newcommand{\sys}[0]{SDO\xspace}
\newcommand{\sysfull}[0]{Self-Defining Operator\xspace}

\definecolor{listinggray}{gray}{0.9}
\definecolor{lbcolor}{rgb}{1.0,1.0,1.0}
\definecolor{lstcomment}{gray}{0.4}
\lstdefinelanguage{golang}{
  keywords={func,return,range,for,if,else,var,continue,break,append,nil},
  keywordstyle=\bfseries,
  comment=[l]{//},
  commentstyle=\itshape\color{lstcomment},
  morestring=[b]",
  stringstyle=\color{black},
  sensitive=true,
}
\lstdefinestyle{gosnippet}{
  language=golang,
  basicstyle=\footnotesize\ttfamily,
  numbers=none,
  showstringspaces=false,
  breaklines=false,
  columns=fullflexible,
  keepspaces=true,
  xleftmargin=2pt,
  aboveskip=2pt,
  belowskip=2pt,
}

\setcopyright{none}

\begin{document}

\date{}

\title{A Self-Defining Operator for Distributed Applications}

\author{Shihang Li, Thomas Anderson, Ratul Mahajan, Simon Peter, and Luke Zettlemoyer\\
  University of Washington}
\renewcommand{\shortauthors}{Li et al.}

\begin{abstract}
To operate distributed applications with high availability, operators must rapidly diagnose and mitigate any incident.
LLM-based agents can help, but unlike experienced human operators they do not improve
as they handle more incidents. Each investigation starts from scratch, re-deriving likely causes and
re-testing hypotheses even when a similar fault was  resolved in the past.
We present a {\em self-defining} operator (\sys) that improves with time by updating its operational behavior and memory.
When \sys resolves an incident, it distills that experience into \emph{playbooks}, which describe reusable diagnostic instructions for future incidents with similar faults, and \emph{detectors}, scripts that take  application telemetry as input and output relevant playbook(s). 
\sys continuously runs the detectors and invokes LLMs only when one fires, so its token cost scales with incident count rather than with operational time.
Across 11~open-source microservice applications, \sys cuts resolution time by
52\% on recurring faults while keeping per-incident cost low.
\sys is open source at \url{https://github.com/self-defining-systems/sdo}.
\end{abstract}

\maketitle

\section{Introduction}
\label{sec:intro}

Distributed applications must remain highly available despite faults in their code,
configuration, dependencies, and infrastructure. When a fault degrades an
application, operators must restore service quickly; the relevant measure is
mean time to resolution (MTTR), the time from incident start until user access to the
application is restored~\cite{sre}. Further, because new features are more likely to cause incidents, resolution speed gates the rate at which new features can be deployed while meeting availability goals.
Thus, fast resolution of operational incidents becomes a key bottleneck to fully realizing the benefits of higher software development productivity brought about by coding agents.

Today, site reliability engineers (SREs) keep applications
healthy~\cite{sre}. What makes an experienced SRE effective is
accumulated operational knowledge: how the application is structured, how it has
failed before, and what resolved each failure. This knowledge is accumulated in runbooks, standard
operating procedures, postmortems, and on-call experience~\cite{sre}. When resolving a new incident,
SREs draw heavily on this knowledge base.
Production incidents are
diverse~\cite{autoarts,fightincidents,cloudstop}, but many recur: a root cause,
such as a misconfigured resource limit or a broken dependency contract, reappears
over time~\cite{flash}, sometimes in new combinations~\cite{autoarts,cloudstop}.
As an SRE operates an application, they learn to recognize these faults and resolve them faster when they recur, driving MTTR down.

AI agents using large language models (LLMs)
have recently emerged as a way to assist SREs. Where
earlier automation ran scripts that applied fixes in static scenarios, an agent can handle incidents dynamically by reasoning over ambiguous telemetry, acting on alerts immediately, and using the same logs, metrics, and APIs as
human operators. Such agents have begun to appear in research~\cite{stratus} and
as commercial cloud services~\cite{awsdevopsagent,azuresreagent}.

However, unlike experienced SREs, today's agents do not accumulate operational knowledge. They do not know when the application is degraded as they cannot distinguish good state from bad; human SREs must trigger them manually when incidents occur or set up other triggering infrastructure. And, after they are triggered, they take a long time to resolve incidents as they must reason from scratch,  re-examining the application structure, re-deriving likely root causes, and retesting hypotheses even when they have already resolved similar incidents.
Consequently, LLM-based agents today are not autonomous, nor do they deliver the MTTR reductions that human SREs achieve from knowledge built up across prior incidents.

We present \sysfull (\sys), an autonomous and learning operator for distributed
applications. It deploys applications from source code and
keeps them healthy as incidents occur. \sys does both by building up \emph{operational memory}, authoring and editing its own application operating procedure as natural-language content and code.
As \sys progresses, it increasingly acts using
memory rather than reasoning from scratch, resolving recurring faults
faster and driving MTTR down as its experience grows.
Every piece of learned behavior is human-readable and editable, so human SREs can audit and extend it if needed.

The key challenge in realizing \sys is building reusable memory from past
deployments and incidents.
To improve over time, \sys must decide what knowledge from each deployment or
incident should persist, how to store it in a form it can use later, and
when that memory applies in future incidents.
This is difficult because the relationship between incidents is not always
visible from their surface symptoms: a later incident may affect different
resources, combine faults from several prior incidents, or
look like an earlier incident at the symptom level despite having a different
root cause. At the same time, because \sys runs continuously,
it must use resources in a cost-effective way.

To make operational memory reusable, \sys defines two artifacts. A
\emph{playbook} records how to recognize and repair a fault, combining
natural-language guidance with scripts for reusable inspection and mitigation steps.
Playbooks are parameterized over roles such as the affected service, missing
configuration object, or failed dependency, so they apply beyond the specific
circumstances of the original incident. Because that parameterization leaves out
application-specific context, \sys also maintains an \emph{architecture
  summary}: a compact description of the application's components and dependencies. The summary grounds a playbook's roles in the
current application and is refreshed as the codebase evolves.

To retrieve memory when it is applicable, \sys writes \emph{detectors}: lightweight non-LLM code that checks over application telemetry. A detector encodes a condition
under which one or more playbooks may apply, such as a violated service
dependency or missing required configuration. When it fires, the detector
returns relevant playbooks and suggests values for playbook parameters,
giving \sys a starting point for investigation.
Detectors give \sys a
cost-effective operating path:
steady-state monitoring is code execution, while LLM token cost is
paid only for deployment configuration changes and incident work.

\sys applies across the application's lifecycle. During initial deployment, \sys
authors the deployment artifacts that bring the application up and defines generic
detectors that monitor error rates for end-user functionality. Once the
application is live, \sys runs a Kubernetes controller for ongoing monitoring and
incident response: a \emph{monitor} evaluates the initial detectors and
invokes the LLM when they report degradation. Thus, before \sys has learned
any fault-specific repair, an incident can still trigger open-ended
investigation. As incidents are resolved, \sys adds fault-specific detectors
that surface matching playbooks. \sys writes detectors as code against a fixed
detector API it cannot modify, and writes playbooks and the architecture summary as
version-controlled files in the application repository.

This paper makes three contributions:
\begin{compactitem}[\labelitemi]
  \item We design \sys, an autonomous operator that brings a Kubernetes
  application from source to production and keeps it healthy by authoring its own
  operating procedure as executable, versioned operational memory: playbooks that
  capture reusable repairs, detectors over live cluster state that surface the
  playbooks that apply, and an architecture summary that models the application.
  \item We implement \sys as a Kubernetes controller centered
  on a monitor that runs agent-authored detectors over live cluster state with
  no LLM on the steady-state path, and a write path that commits all memory to
  the application repository as auditable source.
  \item We evaluate \sys on SREGym~\cite{sregym}, a benchmark for agents that diagnose
  and repair faults in Kubernetes applications. We extend it to support deploying applications from source and to generate new incident variants from its original faults. Across 11~open-source applications, accumulated memory in \sys cuts MTTR by 52\% on recurring faults, including ones that resurface with different symptoms, while detector-based retrieval stays accurate and cheap as memory grows, keeping operating costs low.
\end{compactitem}

\section{Requirements for Autonomous Operators}
\label{sec:motivation}

Operating a production application is continuous work that spans its life:
deploying it, keeping it healthy as faults arise, and scaling and evolving it as the
code or workload changes.
When an incident occurs, operators work through a well-established
sequence~\cite{sre}.
They \emph{detect} the anomaly through alerts, dashboards, or user reports;
\emph{localize} the affected component by narrowing from the whole system
to a service, pod, or configuration file; perform \emph{root-cause
analysis} to determine the underlying cause; and
\emph{mitigate} by applying a fix, such as updating a manifest, restarting
a service, or rolling back a recent change, then verify that the system returns
to health.
This happens under time pressure. Every minute of an outage
interrupts users and erodes trust, and outages remain common and costly
across the industry~\cite{uptimeoutage,autoarts}.

\label{sec:motivation:requirements}

An autonomous operator must work across this full operational lifecycle.
To do so efficiently, it must meet three requirements.
(1) It must own detection rather than wait for an external alerting stack or a human to flag the problem.
Waiting introduces latency and a dependence on external systems.
(2) It must accumulate knowledge across incidents.
Fault classes recur, and linking a new incident to a related past one measurably reduces the effort to resolve it; gaps in this knowledge are a leading cause of delayed mitigation.
(3) It needs to operationalize memory to make it efficient to retrieve when needed. Simply using an LLM to continuously reason about system state is prohibitively expensive given how rarely faults occur relative to uptime. 
Together, these requirements define the gap between an agent that merely responds to incidents and one that operates an application continuously across its life.
We elaborate on these insights in the rest of this section.

\label{sec:motivation:background}

\label{sec:motivation:faults}

\paragraph{Production incidents are diverse but patterned.}
Operating an application is challenging, as production faults are diverse.
Large-scale studies of cloud incidents converge on a broad taxonomy of
root causes: code defects, configuration errors, failed or partial
upgrades, broken dependencies between services, resource exhaustion, and
infrastructure faults~\cite{cloudstop,cloudbugs,autoarts}.
The distribution is dominated by causes outside application code: in one
study of high-severity incidents, roughly 60\% were root-caused to
non-code issues such as service dependencies, infrastructure, deployment,
and configuration~\cite{fightincidents}.
Configuration in particular is a persistent source of
failure, where a single mis-set or missing field can take down a dependent
service~\cite{misconfig,msdebug}.

Failures are also frequently not fail-stop, so a fault can degrade behavior
while the system keeps running and the symptom may be noticed
late~\cite{openstackfaults}.
Effects propagate across service boundaries, so the component where a
problem surfaces is often not the one at fault~\cite{openstackfaults,failatscale}.
Symptoms are often noticed via partial, noisy signals, and the space of plausible
causes is large. Without prior knowledge of the specific application, an
investigator cannot tell which to pursue first~\cite{msdebug}.

Incidents are also not independent draws.
The same root-cause classes reappear throughout multi-year studies of
production outages~\cite{cloudstop,flash,linkedincidents}.
But the relationship between a past incident and a present one is rarely a
surface match: the same root cause can present through different symptoms,
a single incident can combine several causes, and superficially similar
incidents can have unrelated causes~\cite{autoarts,linkedincidents}.
Genuinely novel faults also occur. Hence, prior knowledge works only when it applies without replacing investigation.

\paragraph{Human operators get more effective with experience.}
\label{sec:motivation:practice}
Operators have developed practices to cope with these conditions, and they
center on reusing knowledge from past incidents~\cite{sre,sreworkbook}.
The most direct expression is recognition: an engineer who has seen a
fault class before identifies it from its symptoms and checks the likely
cause directly, skipping the "first principles" investigation a newcomer would
perform~\cite{sre}.
Underlying this is a mental model of the application, including service dependencies, configurations, and past failures.

Other tools and practices come into play as well.
Runbooks codify the mapping from known symptoms to diagnostic steps and
proven fixes. Postmortems convert each incident into
a durable record of what failed and how it was
resolved~\cite{sre,sreworkbook,failatscale}. On-call rotations then spread
this context across a team~\cite{sreworkbook}.
The value is measurable: linking a new incident to related past ones
reduces the effort to mitigate it~\cite{linkedincidents,rcaknowledge},
while gaps in this operational knowledge are a leading cause of delayed mitigation~\cite{fightincidents}. For example, Google reports that a good playbook lowers mean time to repair by roughly 3$\times$~\cite{sre}.

\paragraph{Current SRE agents start incidents cold.}
\label{sec:motivation:agents}
LLM-based agents are a natural fit for this work.
Where earlier automation ran fixed scripts and controllers that handle only
the conditions their authors anticipated, an agent reasons over ambiguous
cluster state and traverses the same diagnostic loop a
human would over the same logs, metrics, and APIs.
Such agents have begun to appear, as research root-cause analysis
assistants~\cite{rcacopilot,rcaagent,flash}, multi-agent diagnosis
systems~\cite{stratus}, and commercial cloud
services~\cite{awsdevopsagent,azuresreagent}.
These systems share a per-incident interaction model: an incident is surfaced, usually by
the existing alerting stack~\cite{opentelemetry,prometheus,pagerduty,grafana,jaeger,datadog,dynatrace,newrelic,splunk,honeycomb,incidentio,zenduty,firehydrant, awsincidentmanager,gcloudops,azurealerts}, and an agent is dispatched to
diagnose and repair the incident. Working context is discarded
afterwards.
Each is, in effect, a freshly hired on-call engineer for a single incident.

\section{\sys Design}\label{sec:design}

\sys is an autonomous operator for distributed applications run via Kubernetes. It learns from the incidents it resolves, and
each resolution updates operational memory, making the next related incident faster to handle.
\S\ref{sec:motivation:requirements} places three requirements on
\sys's design.
First, \sys must \emph{autonomously} operate an application, from deploying it from source to operating it continuously. Second, it must \emph{distill} reusable lessons from each incident,
and it must \emph{retrieve} and apply that lesson when a related fault recurs.
Third, it must do both \emph{at low cost}.

\sys meets these requirements as a long-running agent that \emph{defines itself}. 
It authors code and instructions that determine its runtime
behavior.
In steady state, \sys runs self-authored code that decides when to dispatch \emph{subagents}.
A dispatched subagent uses an LLM to take actions, guided by
instructions authored by earlier invocations, and in turn revises both the
instructions and the dispatching code.

One principle runs through the design: \sys prescribes the interface for
learning, while subagents do the learning.
\sys fixes the form operational memory can take, the interfaces subagents write
against, and what a memory artifact must satisfy before it is committed.

\label{sec:design:overview}

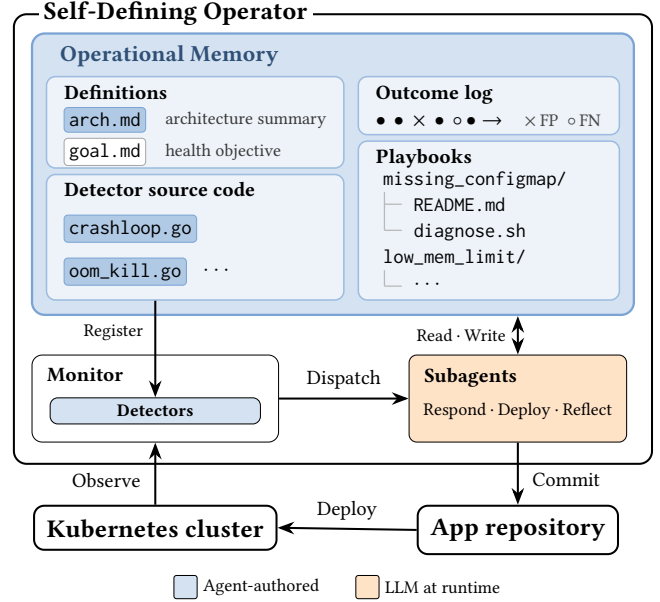
\begin{figure}[t]
  \centering
  \definecolor{authblue}{HTML}{4F86C6}
  \resizebox{\columnwidth}{!}{%
  \begin{tikzpicture}[
    font=\small,
    >=Stealth,
    every node/.style={align=center},
    opbox/.style={draw, thick, rounded corners=5pt, fill=white, inner sep=7pt},
    monbox/.style={draw, rounded corners=4pt, fill=white, inner sep=3pt},
    detbox/.style={draw, rounded corners=3pt, fill=authblue!24, inner sep=3pt},
    rtbox/.style={draw, rounded corners=3pt, fill=orange!22, inner sep=4pt},
    membox/.style={draw=authblue!75, line width=0.9pt, rounded corners=5pt,
      fill=authblue!24, inner sep=4pt},
    card/.style={draw=authblue!55, line width=0.6pt, rounded corners=3.5pt,
      fill=authblue!10, inner sep=0pt},
    cardhdr/.style={anchor=north west, font=\footnotesize\bfseries, inner sep=1pt},
    subhdr/.style={anchor=north west, font=\scriptsize\bfseries, text=gray!60!black, inner sep=1pt},
    fname/.style={draw=authblue!75, rounded corners=1.5pt, fill=authblue!45,
      font=\ttfamily\scriptsize, inner sep=1.5pt, minimum height=2.9mm},
    fnameL/.style={draw=authblue!75, rounded corners=1.5pt, fill=authblue!45,
      font=\ttfamily\footnotesize, inner sep=2pt, minimum height=3.4mm},
    smallbox/.style={draw, thick, rounded corners=4pt, fill=white,
      inner sep=5pt, minimum width=2.0cm, minimum height=0.55cm, font=\bfseries},
    tlabel/.style={anchor=base west, font=\ttfamily\footnotesize},
    note/.style={anchor=base west, font=\scriptsize\itshape, text=gray!60!black},
    swatch/.style={draw, rounded corners=1pt, minimum width=3.2mm,
      minimum height=3.2mm, inner sep=0pt, anchor=west},
    solid/.style={->, thick},
  ]

  \node[opbox, minimum width=8.8cm, minimum height=6.2cm] (sdo) at (0, -0.25) {};
  \node[anchor=west, fill=white, font=\bfseries, inner sep=2pt]
    at ([xshift=10pt]sdo.north west) {Self-Defining Operator};

  \node[membox, minimum width=8.3cm, minimum height=3.88cm, anchor=north] (mem) at (0, 2.6) {};
  \node[anchor=west, font=\small\bfseries, text=authblue!60!black]
    (memtitle) at ([xshift=8pt,yshift=-0.32cm]mem.north west) {Operational Memory};
  \tikzset{humanbox/.style={draw=black!35, rounded corners=1.5pt, fill=white,
    font=\ttfamily\footnotesize, inner sep=2pt, minimum height=3.4mm}}

  \node[card, minimum width=4.10cm, minimum height=1.26cm, anchor=north] (defcard) at (-1.90, 2.0) {};
  \node[card, minimum width=4.10cm, minimum height=1.73cm, anchor=north] (detcard) at (-1.90, 0.65) {};
  \node[card, minimum width=3.60cm, minimum height=0.80cm, anchor=north] (outcard) at (2.15, 2.0) {};
  \node[card, minimum width=3.60cm, minimum height=2.19cm, anchor=north] (pbcard) at (2.15, 1.11) {};

  \node[cardhdr] (hdef) at ([xshift=6pt,yshift=-2pt]defcard.north west) {Definitions};
  \node[fnameL, anchor=north west] (archf) at ([xshift=1pt,yshift=-2pt]hdef.south west) {arch.md};
  \node[anchor=west, font=\scriptsize, text=gray!45!black] at ([xshift=4pt]archf.east) {architecture summary};
  \node[humanbox, anchor=north west] (goalf) at ([yshift=-2pt]archf.south west) {goal.md};
  \node[anchor=west, font=\scriptsize, text=gray!45!black] at ([xshift=4pt]goalf.east) {health objective};

  \node[cardhdr] (hout) at ([xshift=6pt,yshift=-2pt]outcard.north west) {Outcome log};
  \node[anchor=north west, font=\footnotesize, inner sep=1pt] (outm)
    at ([yshift=-2pt]hout.south west) {$\bullet\,\bullet\,\times\,\bullet\,\circ\,\bullet\rightarrow$};
  \node[anchor=west, font=\scriptsize, text=gray!55!black] at ([xshift=3.5pt]outm.east) {$\times$\,FP\ \ $\circ$\,FN};

  \node[cardhdr] (hdet) at ([xshift=6pt,yshift=-2pt]detcard.north west) {Detector source code};
  \node[fnameL, anchor=north west] (detf1) at ([xshift=1pt,yshift=-5.8pt]hdet.south west) {crashloop.go};
  \node[fnameL, anchor=north west] (detf2) at ([yshift=-5.8pt]detf1.south west) {oom\_kill.go};
  \node[anchor=west, font=\footnotesize, inner sep=1pt]
    at ([xshift=5pt]detf2.east) {$\cdots$};

  \node[cardhdr] (hpb) at ([xshift=6pt,yshift=-2pt]pbcard.north west) {Playbooks};
  \node[anchor=base west, font=\ttfamily\footnotesize] (pf1) at (0.57, 0.50) {missing\_configmap/};
  \node[anchor=base west, font=\ttfamily\footnotesize] (pr1) at (0.99, 0.14) {README.md};
  \node[anchor=base west, font=\ttfamily\footnotesize] (ps1) at (0.99, -0.22) {diagnose.sh};
  \node[anchor=base west, font=\ttfamily\footnotesize] (pf2) at (0.57, -0.58) {low\_mem\_limit/};
  \node[anchor=base west, font=\footnotesize] (pe2) at (0.99, -0.94) {$\cdots$};
  \begin{scope}[gray!55, line width=0.5pt]
    \draw (0.71, 0.40) -- (0.71, -0.13);
    \draw (0.71, 0.23) -- (0.95, 0.23);
    \draw (0.71, -0.13) -- (0.95, -0.13);
    \draw (0.71, -0.68) -- (0.71, -0.90);
    \draw (0.71, -0.90) -- (0.95, -0.90);
  \end{scope}

  \node[monbox, minimum width=3.4cm, minimum height=1.2cm] (monitor) at (-2.45, -2.45) {};
  \node[anchor=north west, font=\footnotesize\bfseries, inner sep=2pt]
    at ([xshift=4pt,yshift=-3pt]monitor.north west) {Monitor};
  \node[detbox, minimum width=2.9cm, minimum height=0.36cm] (det)
    at ([yshift=-0.17cm]monitor.center) {};
  \node[font=\scriptsize\bfseries] at (det.center) {Detectors};

  \node[rtbox, minimum width=3.0cm, minimum height=1.2cm] (resp) at (2.55, -2.45) {};
  \node[anchor=north west, font=\footnotesize\bfseries, inner sep=2pt]
    (subt) at ([xshift=4pt,yshift=-3pt]resp.north west) {Subagents};
  \node[font=\scriptsize] at ([yshift=-0.17cm]resp.center) {Respond\,$\cdot$\,Deploy\,$\cdot$\,Reflect};

  \draw[solid] (monitor.east |- resp) -- (resp.west)
    node[midway, above, font=\footnotesize] {Dispatch};

  \draw[solid] (det.north |- detcard.south) -- (det.north)
    node[pos=0.33, left=1pt, font=\scriptsize] {Register};
  \draw[<->, thick] (resp.north) -- (resp.north |- mem.south)
    node[midway, left=1pt, font=\scriptsize] {Read\,$\cdot$\,Write};

  \node[smallbox, anchor=north] (cluster) at ([xshift=-2.45cm, yshift=-0.55cm]sdo.south) {Kubernetes cluster};
  \node[smallbox, anchor=north] (repo) at ([xshift=2.55cm, yshift=-0.55cm]sdo.south) {App repository};

  \draw[solid] (resp.south) -- (repo.north)
    node[pos=0.6, right=2pt, font=\footnotesize] {Commit};
  \draw[solid] (repo.west) -- (cluster.east)
    node[midway, above, font=\footnotesize] {Deploy};
  \draw[solid] (cluster.north) -- (monitor.south)
    node[pos=0.4, left=2pt, font=\footnotesize] {Observe};

  \node[swatch, fill=authblue!24] (sw2) at (-2.2, -5.05) {};
  \node[anchor=west, font=\scriptsize, inner sep=2pt] (l2) at (sw2.east) {Agent-authored};
  \node[swatch, fill=orange!22] (sw3) at ([xshift=12pt]l2.east) {};
  \node[anchor=west, font=\scriptsize, inner sep=2pt] (l3) at (sw3.east) {LLM at runtime};

  \end{tikzpicture}}%
  \caption{\sys is an \emph{autonomous} operator. It brings a Kubernetes
  application up from its source repository and keeps it healthy. Its
  agent-authored \emph{operational memory} (detectors, playbooks, architecture
  summary, and an outcome log) is versioned alongside the application source. The
  \emph{Monitor} runs detectors that dispatch \emph{subagents}
  to resolve incidents.}
  \label{fig:architecture}
\end{figure}

\paragraph{Overview.}
Figure~\ref{fig:architecture} shows an overview of \sys.
When the application is not yet running, a human-provided deployment objective brings it from
source to a healthy deployment (\S\ref{sec:design:deploy}).
Once it is live, a \emph{monitor} watches application health and dispatches
a \emph{responder} subagent as incidents arise (\S\ref{sec:design:learning}).
After resolving an incident, the responder \emph{reflects},
by distilling what the session learned into operational
memory. A code commit deploys the resolution, while refreshing memory so it stays aligned with the application (\S\ref{sec:design:upkeep}).

\sys's \emph{operational memory}, the knowledge it accumulates, has five kinds of artifacts: \emph{playbooks}, \emph{detectors}, an \emph{outcome log}, an \emph{architecture summary}, and a \emph{health objective}.
The architecture summary captures the application's logical structure, such as
its services and the dependencies between them, while the health objective captures human-provided input defining healthy operations for the application (\S\ref{sec:design:deploy}).
A playbook contains reusable repair instructions generalized from a resolved incident (\S\ref{sec:design:learning}).
A detector is non-LLM, runnable code over cluster state that quickly detects a fault and
surfaces the playbooks relevant to addressing it (\S\ref{sec:design:retrieval}).
For example, a detector might flag any deployment that references a missing Kubernetes ConfigMap.
Finally, the outcome log marks each incident response as a success, a false positive (FP), or a false negative (FN), used by \sys to refine its detectors (\S\ref{sec:design:upkeep}).
Memory artifacts are maintained as durable text files, enabling human operators to read, edit, or seed any artifact.

\subsection{Application Deployment}
\label{sec:design:deploy}

Initial deployment is where \sys first populates operational memory.
The human operator gives \sys a natural-language health objective: a
definition of what it means for the application to be healthy, the public-facing
endpoints it must expose, and any other requirements (e.g., which services to co-locate).
From this objective, \sys brings the application from source to a running state
that satisfies it.
It does this with two subagents: a
\emph{deployer} produces the artifacts needed to run the application together with an architecture summary that models it, and
a \emph{health judge} checks whether the
application is healthy.
The two subagents iterate: the deployer brings the application up, the judge decides
whether it is healthy, and the deployer revises until it is.
Bringing an application up takes many iterations, enough that the detail
accumulated across them would overrun a single context window;
\sys therefore starts each role with a fresh context on every handoff and
carries continuity only through the architecture summary and objectives.

\paragraph{The deployer.}
The deployer produces the architecture summary, capturing the application's
main components, their relationships and dependencies, their configuration, and the build graph.
The summary is small by construction and orients the deployer toward a working deployment without re-reading the source across attempts.
The deployer commits each attempt, so a later subagent session, started fresh, recovers from the summary and its own committed history.
From this understanding, it authors the deployment artifacts such as Kubernetes manifests and brings the application up.

\paragraph{The health judge.}
The health judge uses the human operator's definition of what it
means for the application to be healthy to author programs that decide whether the application
has reached a healthy state: programmatic checks over cluster and application-level state. The judge then
identifies failure patterns the checks would miss and revises
them to handle those patterns, iterating until the checks are robust.
These checks are \sys's first detectors: they measure
application health rather than look for any specific fault.
During deployment they are the success signal. The application is up when they
all report healthy, and the deployer and judge iterate until they do.
Once the application is live, they become a safety net. Any fault
that manifests in observable state, whether as degraded user-facing behavior or
as divergence from the application's desired state, trips one of them and dispatches a responder.
Because they are anchored in the human operator's objectives and authored by a role separate from the responder that repairs
incidents, they remain an independent measure of health that the responder
cannot redefine to declare premature victory.

\subsection{Response to New Incidents}
\label{sec:design:learning}

Once the application is live in production, incidents may
arise. The monitor dispatches a responder when a detector fires
(\S\ref{sec:design:retrieval} describes how \sys detects incidents and dispatches responders).
The generic health detectors authored during deployment (\S\ref{sec:design:deploy}) are always present, so any fault that degrades application health dispatches a responder even before a fault-specific detector exists.
The human operator can also dispatch a responder directly for a fault the health detectors miss; \sys logs the miss as a false negative and authors a detector for it (\S\ref{sec:design:upkeep}).
The responder's measure of success is every detector returning to a healthy report; the responder works until no detector fires.

\paragraph{Quick playbook check.}
The responder browses the playbook library first.
Browsing is tiered so it stays cheap as memory grows. \sys keeps an index that
names each playbook with a one-line description of the fault it repairs. The
responder reads this index first, opening a playbook's guidance only when its index
entry looks relevant and descending into its scripts only when that guidance
fits the current cluster. The responder thus scans the whole library without
reading every playbook in full. 

\paragraph{Hypothesis-driven investigation.}
If no playbook is found (the default upon first deployment), the responder
investigates from live cluster state with an iterative hypothesis-revision loop.
Each round forms a candidate cause from the current symptoms and the
architecture summary, then tests it: if this cause were the fault, the cluster
would exhibit some observable consequence, which the responder checks for,
confirming the cause or eliminating it and forming a revised hypothesis. Once a cause is confirmed, the responder mitigates it.
Because a complex incident can stem from several contributing
causes, the responder keeps forming, testing, and mitigating until every
detector reports healthy. The architecture summary seeded at deployment
(\S\ref{sec:design:deploy}) bounds the candidate causes to those consistent with
the application's structure.
The responder can spawn subagents to pursue competing
hypotheses concurrently, each returning only its conclusion so the parent's
context stays focused.

When the responder needs to check a property of the cluster, it authors a parameterized script, runs it, and keeps the
script if it proves useful. This is similar to how SREs accumulate shell scripts
across incidents.
Suspecting a configuration problem, for instance, the responder could write a script
to enumerate a service's ConfigMap references and report missing targets,
run it against the crash-looping service, and confirm the missing reference.
When it later reflects, the responder keeps successful scripts as part of the playbook.

\paragraph{Post-fix reflection.}
A successful session ends with a \emph{reflect} step.
Because the responder that just handled the incident does the reflection, its
decisions are still in context, and it abstracts the resolution directly.
As it reflects, the responder distills the resolution into a playbook and authors a
detector that can flag this fault going forward.
The responder commits the new playbook and detector to the repository, and a redeploy
registers the detector with the monitor.
\sys only commits valid detectors: a detector must
compile against the \sys-prescribed interface, and each playbook it names must exist.
From this point the fault is covered: a recurrence trips the detector rather
than requiring another open-ended investigation, the monitor surfaces the
relevant playbooks, and the responder decides whether each one applies to the
current incident.

\subsection{Fast Response to Known Incidents}
\label{sec:design:retrieval}

\begin{listing}[t]
\begin{lstlisting}[style=gosnippet, basicstyle=\scriptsize\ttfamily]
// detectors/missing_configmap.go
// a pure function that detects missing config map entry
func Detect(c Cluster) []Finding {
  var out []Finding
  for _, d := range c.Deployments() {
    for _, cm := range d.ConfigMapRefs() {
      // fault: the referenced ConfigMap is absent
      if c.ConfigMaps().Has(cm) { continue }
      // surface the playbook, binding its parameters
      // to the flagged deployment and ConfigMap
      out = append(out, Finding{
        Playbook: "missing_configmap",
        Bind:     Params{"service": d.Name, "configMap": cm},
      })
    }
  }
  return out
}
\end{lstlisting}
\caption{The detector for the missing-ConfigMap fault. It surfaces the
\texttt{missing\_configmap} playbook with its parameters bound to the
deployment and ConfigMap it flags.}
\label{fig:detector}
\end{listing}

\paragraph{Detectors flag incidents, surface playbooks.}
Given a known incident, a corresponding detector surfaces relevant playbooks.
To do so, a detector declares two things: the cluster state it reads, which tells the
monitor when to re-run the detector, and the playbooks it may surface.
Its logic decides, from the state it reads, whether its condition holds and
which of its playbooks to surface. To help the responder, a detector may also localize the condition by naming the offending object and any related objects.
In other words, detectors are pure functions over cluster state, using accessors to the application's services, pods,
deployments, ConfigMaps, endpoints, and events (see Listing~\ref{fig:detector} for an example).
A detector reads cluster state by default. For a signal that is not in
cluster state, such as an application-level latency metric, a reflecting responder may
author a telemetry, integrated with standard observability tooling~\cite{prometheus,opentelemetry}, that feeds the signal into the queryable view, so the
detector surface grows with the application.

\paragraph{Detectors are cheap and robust.}
Because a detector is deterministic code over structured state, the monitor
evaluates it on its own schedule without any LLM call, making it cheap. Steady-state cost is therefore detector evaluation alone. LLM token cost scales
with the number of incidents rather than with operating time.

Only a detector that fires persistently across evaluations dispatches a responder,
with the recent evaluation history and the conditions it flagged provided as
initial context. A transient failure that clears on the next evaluation is treated as a fluke.
Detector evaluation is deterministic: the same cluster state always
yields the same result.
In our example (Listing~\ref{fig:detector}), when the fault recurs on a different deployment, the detector
names the deployment and the absent ConfigMap, and the responder binds the
playbook's parameters to them, mapping each parameter to the resource the
detector flagged.

\paragraph{Playbooks guide from root cause to resolution.}
The responder confirms the surfaced playbooks against current state. Once they are confirmed, it runs the relevant playbooks to yield a resolution. Playbooks have two components.
The first is a pair of natural-language instructions: how to decide whether
the current cluster has the root cause this playbook describes, and, if it
does, how to repair it.
The second is a set of shell scripts the responder can run to
inspect the live cluster to help it determine the root cause.
Both the instructions and scripts use placeholder names rather than literal
resource names, so a playbook applies to any later incident with the same
root cause regardless of which specific services or configuration objects are involved.
Reflection chooses what to parameterize.

\paragraph{Fixes are double-checked.}
The session ends when detectors report the application healthy again.
In this case, the responder commits the repair as a source change and redeploys.
Confirmation guards the repair: the responder re-checks the condition the
playbook assumes, here that the named ConfigMap is indeed absent, before
acting, and if the condition does not hold, the
session falls back to the open-ended investigation of
\S\ref{sec:design:learning}.
When several faults coincide, say a missing ConfigMap on one service and a port
misconfiguration on another, several detectors fire at once, and the monitor
batches them into one session in which the responder applies each surfaced repair,
bound to its own resources.

\subsection{Refining Memory Over Time}
\label{sec:design:upkeep}

Applications change over time. Hence, \sys evolves its memory to
keep it aligned with the live application. To do so, it refines detectors based on incident outcomes and it revises the architecture summary with new repository commits.

\paragraph{Refining from outcomes.}
A detector can map to its playbooks too broadly or too narrowly, and \sys has
a signal for each.
A detector is too broad when it fires but the responder, on inspection, finds its
condition absent and applies no playbook. For example, the missing-ConfigMap
detector might flag a reference marked optional, which is not a fault.
It is too narrow when the responder resolves a fault with a playbook that the detector failed to surface, reaching it only by open-ended
investigation.

\sys records both signals: for each session, \sys logs which detectors fired and
which playbooks the responder applied, and commits this append-only outcome log to operational memory.
The responder reads this log when it reflects and may adjust the mapping, tightening
or splitting a detector that fired too broadly and widening one that fired too
narrowly.
Over successive incidents this sharpens the detector-to-playbook mapping.
The same log grounds structural edits: detectors that fire together across
sessions suggest one root cause to merge, and a detector that leads to
divergent repairs suggests two to split.

\paragraph{Keeping memory aligned.}
Each architecture summary is tagged with the Git commit at which it was
generated. On a later commit, \sys passes the diff against that commit to the
responder at its next reflection, which classifies whether the change is topology-relevant
(services added or removed, dependency edges changed, configuration keys
renamed) and regenerates the affected portion of the summary if so.
Playbooks and detectors whose referenced entities the diff removed or renamed
are marked for review, which the agent revisits at its next reflection.

\sys does not retire memory by disuse.
Operational knowledge has a long-tail value distribution: a playbook that
applies once a year is precisely the entry a responder could not reason about from
first principles, yet usage-based pruning (LRU, frequency thresholds) would
retire exactly these rarely-used entries.
\sys refines an entry that proves wrong, through the false-positive and
false-negative signals above, and retains an entry that stays quiet.
Memory changes only through edits the agent makes during reflection.

\section{Implementation}\label{sec:implementation}

\sys is split into an agent runtime and a controller.
The agent runtime is approximately 6{,}000~lines of Python; the monitor and
detector runtime is a Kubernetes controller written in approximately 3{,}000~lines of Go.
The agent runtime exposes a generic interface that decouples \sys from any
single agent implementation.
We provide two backends: our own harness built on the
Pydantic~AI agent framework~\cite{pydanticai}, and a portable adapter that
drives off-the-shelf coding agents such as Claude Code, Codex, and
OpenCode.
The agent has standard coding-agent tools, a shell and file read/write
primitives, and it may spawn nested agents within a session.
No agent carries state between sessions.

\paragraph{Standardized agent interactions.}
\sys standardizes the interactions it requires of an agent through
Model Context Protocol (MCP)~\cite{mcp}
servers, so any backend behind the agent interface participates
identically.
The agent reports the structured outcomes \sys depends on through these
servers, for instance recording that it has reached a diagnosis.
\sys also exposes CLI tools the agent may run while editing memory, such
as checking that a detector compiles and that it only references previously saved playbooks.
The agent invokes these checks itself, and \sys guards the commit path to
reject any change that fails them, so a session cannot persist memory that
breaks these validity checks.

\paragraph{Monitor and detectors.}
The monitor runs in the controller and reads cluster state through the
Kubernetes API.
Each detector is compiled into the controller and loaded from the
repository, and it declares the resource kinds it reads and
the interval at which it runs (\S\ref{sec:design:retrieval}).
The monitor evaluates each detector on its interval, snapshotting a read-only,
typed view of the cluster state and passing it to the detector
(the \texttt{Cluster} value in
Listing~\ref{fig:detector}).

\paragraph{Externalizing memory.}
\sys stores its memory in the application's Git repository: the
architecture summary and playbooks as Markdown (with each playbook referencing its
scripts by path), the scripts as their own files, and detectors as Go
source.
The agent reads and writes this tree with file primitives and commits
changes with git; a redeploy compiles and registers any new or changed
detectors with the monitor.
We chose a plain file tree over an embedding store: the rapid adoption of
agent skills~\cite{agentskills}, which organize domain knowledge as
file-tree hierarchies, shows that such layouts are effective for LLM
retrieval while remaining human-readable and version-controllable.
Keeping memory in the repository also makes every change to the operator's
behavior an auditable commit.

\section{Evaluation}
\label{sec:eval}

We evaluate \sys to answer four questions:
\begin{compactenum}[1.]
\item Do the speed and effectiveness of end-to-end incident response improve as \sys accumulates
  memory?
  (\S\ref{sec:eval:effectiveness})
\item Can \sys deploy applications reliably and efficiently? (\S\ref{sec:eval:deployment})
\item Does learned operational memory improve resolution of unseen recurrences without
  degrading performance on novel faults?
  (\S\ref{sec:eval:memory-transfer})
\item Do detectors retrieve
  relevant operational memory accurately and cheaply?
  (\S\ref{sec:eval:retrieval})
\end{compactenum}

\paragraph{Benchmark.}
We evaluate incident resolution on a variant of SREGym~\cite{sregym},
a benchmark of 90~fault-injection problems spanning distinct root causes
across multiple microservice applications deployed on Kubernetes.
Each scenario introduces a fault, such as a misconfiguration, resource
constraint, or dependency failure, into one or multiple running applications.
It measures whether an agent can \emph{diagnose} the root cause and
\emph{mitigate} it by resolving the fault. By design, SREGym starts each incident scenario from scratch, by redeploying the involved applications. However, \sys may retain operational memory across scenarios.

Across the evaluation, we use 11~open-source microservice applications
(Table~\ref{tab:eval-apps}).
The incident-resolution experiments apply fault scenarios derived from
SREGym's problem suite to four of these applications (marked in
Table~\ref{tab:eval-apps}); the deployment experiment uses all 11
applications to stress deployment from source.

\begin{table}[t]
  \centering
  \footnotesize
  \caption{Evaluated applications. Checkmarks indicate the applications
  used in the incident-resolution experiments; the deployment experiment
  uses all 11.}
  \label{tab:eval-apps}
  \setlength{\tabcolsep}{2pt}
  \begin{tabular}{@{}p{0.30\columnwidth}ccp{0.47\columnwidth}@{}}
    \hline
    Application & Incidents & Svcs & Language(s) \\
    \hline
    hotelReservation~\cite{deathstarbench} & \checkmark & 10 & Go \\
    media~\cite{deathstarbench} & & 14 & C++ \\
    socialNetwork~\cite{deathstarbench} & \checkmark & 13 & C++ \\
    fleetcast~\cite{fleetcast} & \checkmark & 2 & Python, JavaScript \\
    onlineboutique~\cite{onlineboutique} & & 11 & Go, Java, Python, JavaScript, C\# \\
    pitstop~\cite{pitstop} & & 13 & C\#, JavaScript \\
    sockshop~\cite{sockshop} & & 8 & Java, Go, JavaScript, C\# \\
    sockshop-helidon~\cite{sockshophelidon} & & 6 & Java \\
    teastore~\cite{teastore} & & 7 & Java \\
    teastore-saga~\cite{teastoresaga} & & 9 & Java \\
    train-ticket~\cite{trainticket} & \checkmark & 41 & Java \\
    \hline
  \end{tabular}
\end{table}

We extend SREGym in two ways to create more diverse fault scenarios.
First, we analyze the 90 faults in SREGym and use them to
create fault templates, which permute an original SREGym problem along
a common set of axes: the application and service(s)
that are impacted by the fault, and one or more fault-specific parameters
(\eg port numbers, environment variable names and values).
Using these fault templates, we generate different variants of the
same fault, where each fault impacts the deployed application in
a different way, depending on the fault parameters that are used.
Second, we compose multiple faults into a single incident.
We create an incident stream generator that creates new incidents
by composing one or more faults generated from the above templates.
The incident stream is computed deterministically given the same
seed.

\paragraph{Agent harness.}
Across all experiments, every agent is instantiated with Claude Code
using Claude Sonnet 4.6.
Baselines differ only in the memory available to the agent and in the
mechanism used to retrieve that memory; unless stated otherwise, they
use the same model, tools, Kubernetes access, and benchmark interface.

\subsection{Incident Resolution Effectiveness}
\label{sec:eval:effectiveness}

We first investigate whether \sys improves end-to-end incident response
as it accumulates fault-specific operational memory.
To do so, we evaluate the aggregate learning curve over a
deterministic stream of 200 incidents, where each incident may involve one
or multiple of the four fault-target applications (Table~\ref{tab:eval-apps}).
Each incident deploys its target applications, injects its composed
faults, and tears the applications down after resolution.
We replay the identical stream for every configuration.
\sys builds operational memory per application in each application repository.
Each agent is given repositories of the applications involved
in the incident.

\paragraph{Metrics.}
We measure incident resolution effectiveness via the following metrics.
\emph{Time to resolution} (TTR) is the end-to-end time from incident
start to resolution, where the incident starts when SREGym injects
the faults into the application, and is marked resolved when the agent
submits a benchmark-specific tool call indicating resolution.
\emph{Solve rates} report diagnosis success, mitigation success, and
end-to-end success.
Diagnosis success tests whether the agent identified the correct root
causes for the faults; mitigation success tests whether the agent
remediated the faults; end-to-end success requires both.
Token usage is the total number of LLM tokens consumed while resolving
each incident.

\paragraph{Baselines and configurations.}
We compare four configurations.
\emph{Memoryless} is Claude Code with no persistent incident memory,
matching the best-performing SREGym baseline~\cite{sregym}.
\emph{History-Embed} and \emph{History-LLM} are baseline memory systems
that store prior incident histories.
Each incident is stored in two layers: the full trajectory of the agent's incident
response and a short summary generated by the agent.
History-Embed retrieves histories by a popular embedding and vector retrieval configuration: we use BAAI/bge-base-en-v1.5~\cite{bge_embedding} to compute each incident summary's embedding and Qdrant~\cite{qdrant} for document indexing and retrieval, returning the top five histories by similarity.
History-LLM mimics an emerging pattern of retrieving knowledge in agent context (\eg agent skills~\cite{agentskills}): the agent directly reads the summaries and chooses histories
that are relevant to the current incident.
In both variants, the agent can grep over retrieved trajectories.

\begin{figure*}[t]
  \centering
  \includegraphics[width=\textwidth]{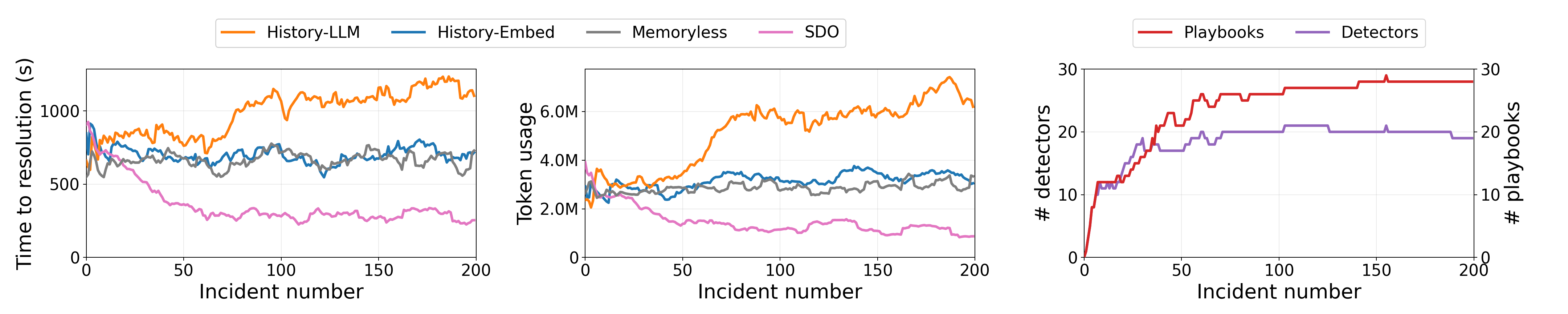}
  \caption{Timeline metrics versus cumulative incident number for
  Memoryless, History-Embed, History-LLM, and \sys.
  \emph{Left} and \emph{center} report the rolling average (window=30)
  of time to resolution and token usage, showing whether each memory system
  improves as its incident history grows.
  \emph{Right} reports the growth of \sys's operational memory, measured by the
  number of detectors and playbooks, over the same incident stream.}
  \label{fig:incident-timelines}
\end{figure*}

\begin{figure}[t]
  \centering
  \begin{minipage}[t]{0.49\columnwidth}
    \centering
    \includegraphics[width=\linewidth]{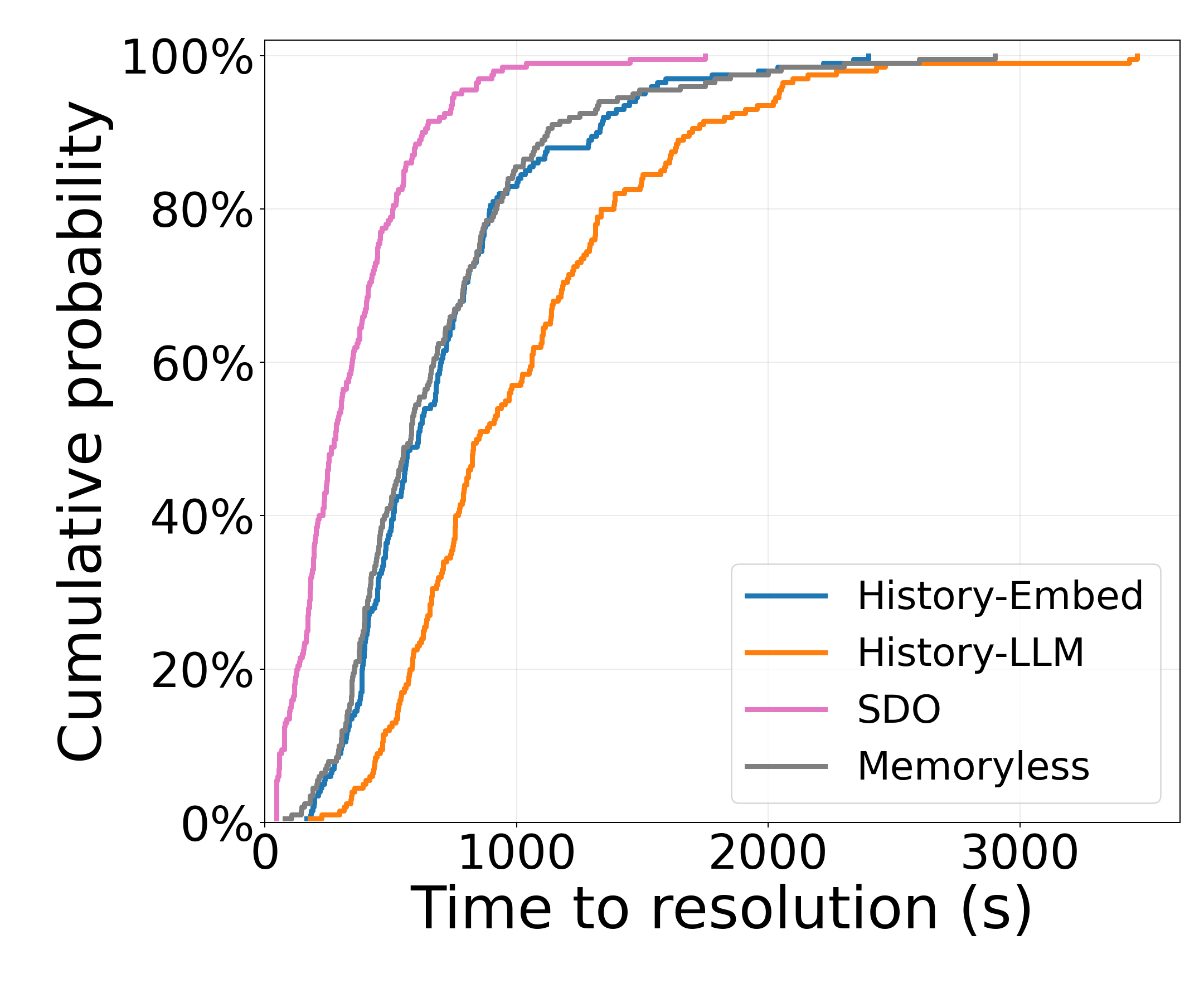}
    \small (a) Time to resolution.
  \end{minipage}\hfill
  \begin{minipage}[t]{0.49\columnwidth}
    \centering
    \includegraphics[width=\linewidth]{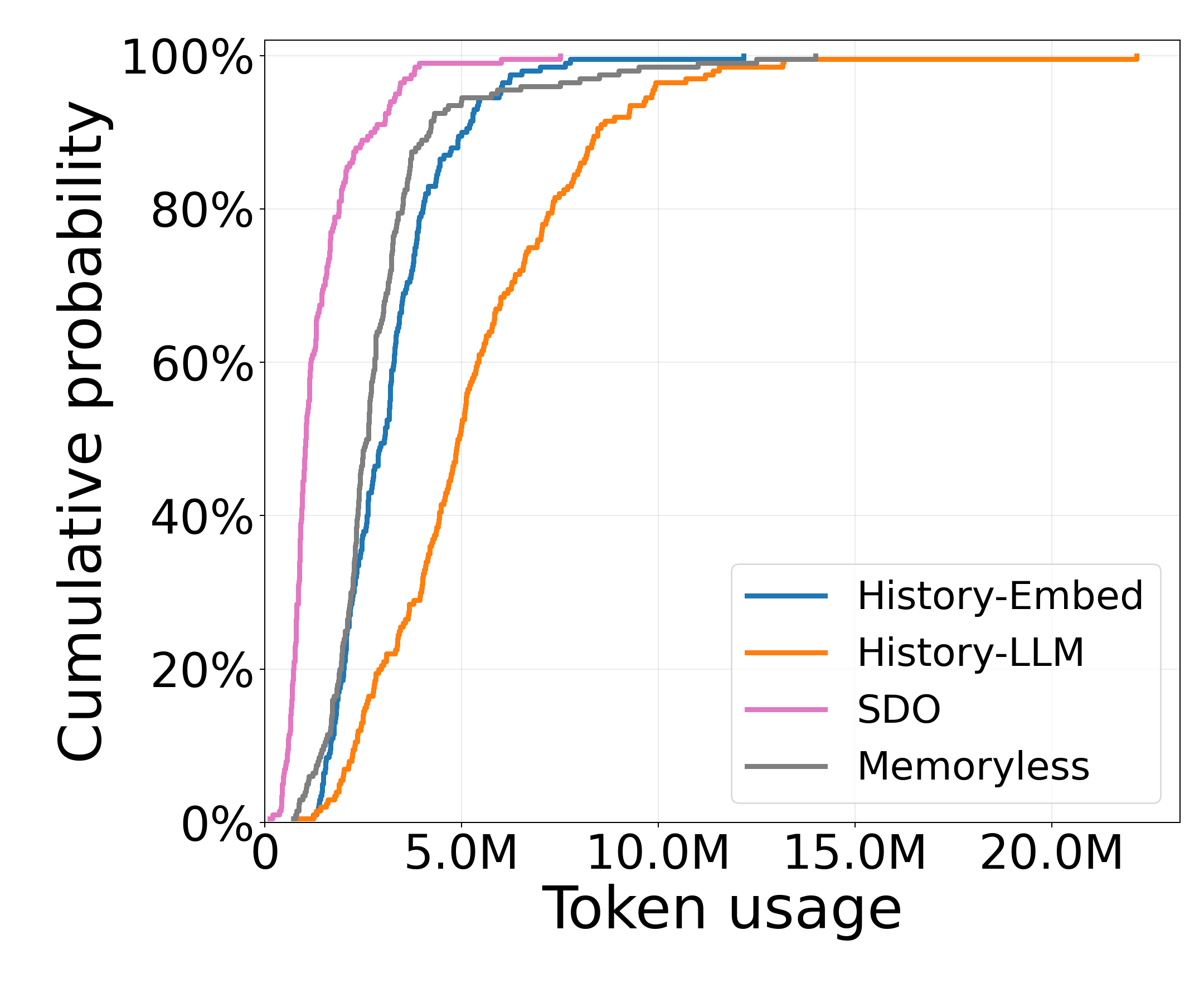}
    \small (b) Token usage.
  \end{minipage}
  \caption{CDFs of time to resolution and token usage across the
  sequential incident-resolution experiment.}
  \label{fig:ttr-token-cdfs}
\end{figure}

\begin{table}[t]
  \centering
  \footnotesize
  \caption{Agent incident solve rates.}
  \label{tab:incident-solve-rates}
  \begin{tabular*}{\columnwidth}{l@{\extracolsep{\fill}}ccc}
    \hline
    Configuration & Diagnosis & Mitigation & End-to-end \\
    \hline
    Memoryless & 72.3\% & 78.2\% & 62.7\% \\
    History-Embed & 65.3\% & 76.6\% & 59.4\% \\
    History-LLM & 43.4\% & 63.3\% & 31.8\% \\
    \sys & 75.3\% & 80.9\% & 65.4\% \\
    \hline
  \end{tabular*}
\end{table}

\paragraph{Results.}
All systems see substantial per-incident variance because the incident
generator composes different numbers and kinds of faults into each
problem. We compare configurations on the same incident sequence and
focus on trends and distributions rather than individual incidents.
Figures~\ref{fig:incident-timelines} and~\ref{fig:ttr-token-cdfs}
report rolling-average timelines (window of 30 incidents) and distributions of TTRs and token usage, and
Table~\ref{tab:incident-solve-rates} reports solve rates for the
entire incident stream. We find that \sys resolves incidents faster and cheaper as it
accumulates memory.
Early in the stream, \sys behaves like the baselines because it has
little fault-specific memory.
As detectors and playbooks accumulate, the rolling TTR drops below all
baselines, and token usage falls with it.

The improvement comes from two coupled mechanisms: detectors reduce
search, and playbooks reduce resolution work.
First, detectors retrieve the relevant memory cheaply and localize it to
the resources implicated by current cluster state.
Second, the playbook turns prior work into reusable procedure.
The agent still reasons over the live cluster and can reject a misleading
playbook when evidence contradicts it, but a correct playbook shifts much
of the work from generating new shell commands and hypotheses to invoking
agent-authored scripts with the right parameters.

The right panel of Figure~\ref{fig:incident-timelines} shows how \sys's memory
grows over the stream.
Early incidents add playbooks and detectors rapidly: with little memory to draw
on, \sys investigates from scratch and distills each resolution into new memory,
sometimes adding several playbooks and detectors at once when an incident exposes
independent faults.
Detectors stay fewer than playbooks because they key on symptoms while playbooks
key on root causes, and distinct root causes can manifest the same symptoms.
Over incidents, \sys keeps
improving the entries it already has, refining playbooks and merging redundant
ones.
Inspection of the repository git history shows this consolidation: agents
repeatedly refactored command sequences they had run as sequential shell calls
into reusable playbook scripts, which grew into broader utility tools with
subcommands and flags for common diagnostic and mitigation tasks.
Agents revised the playbook text to call these tools directly; a later entry,
for example, runs one subcommand to test connectivity and another to enumerate
affected containers on a dependency failure.
In effect, \sys builds a small suite of operational CLI tools that speed its
interaction with the cluster.

Memoryless shows the cost of having no accumulated operational state.
Memoryless has high variance from problem difficulty, but its rolling TTR
and token usage are flat. It carries no state between incidents, so
it repeatedly pays the same orientation and investigation cost.

History-Embed shows that cheap retrieval is not enough when memory is
stored as raw incident history.
History-Embed has TTR and token profiles similar to Memoryless, but a
lower solve rate.
Its retrieval operates over raw incident histories, so symptom-similar
incidents can outrank incidents with the same root cause, and many
near-duplicate past incidents reduce retrieval diversity as memory grows.
When the retrieved summary does not match current cluster state, the
agent often ignores it and falls back to scratch investigation.
Its solve rate is lower than Memoryless because retrieval by symptom similarity sometimes
returns irrelevant incidents, misleading the agent.

History-LLM shows another failure mode: active retrieval over
unstructured memory can take much longer than no memory.
Because one raw history is added after every incident, the search space
grows linearly with the stream.
As the number of past incidents increases, the agent spends more tokens reading summaries and grepping over potentially relevant
trajectories, and more time deciding whether prior incidents apply.
Claude Code can spawn nested agents in all configurations, but we observe
much more nested-agent use in History-LLM because
retrieval itself is delegated to model reasoning.
Those nested agents often duplicate relevance checks and Kubernetes probes
across similar prior incidents, and the main agent frequently
double-checks their findings by reading the flagged trajectories again, incurring high token usage.
Trajectory inspection shows an anchoring effect: once an early retrieved
history appears relevant, the main agent tends to investigate that
history's hypothesis disproportionately, even when later evidence points
elsewhere.
It increases context, time, and cost, while also lowering diagnosis and
mitigation accuracy.

\begin{figure*}[t]
  \centering
  \begin{subfigure}[t]{0.49\textwidth}
    \centering
    \includegraphics[width=\linewidth]{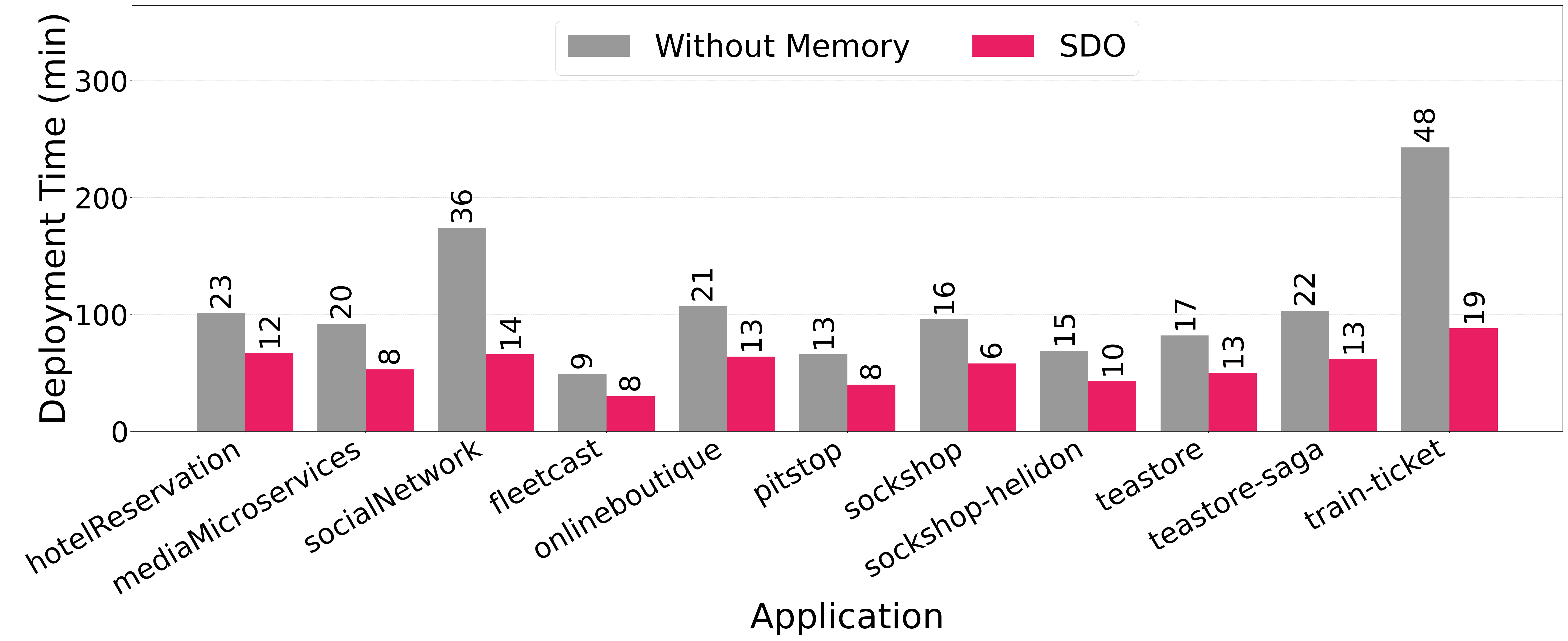}
    \caption{Deployment time; bar labels show iteration count.}
    \label{fig:deploy-time}
  \end{subfigure}\hfill
  \begin{subfigure}[t]{0.49\textwidth}
    \centering
    \includegraphics[width=\linewidth]{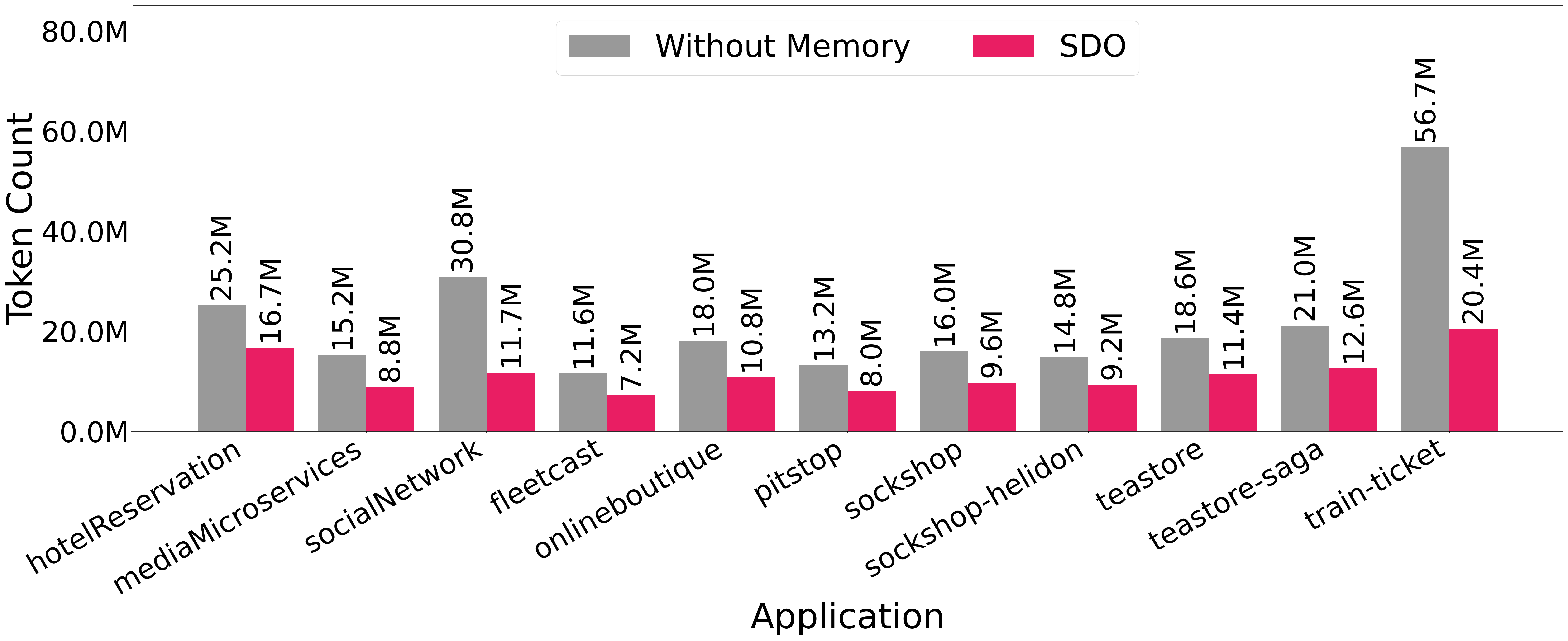}
    \caption{Token usage.}
    \label{fig:deploy-tokens}
  \end{subfigure}
  \caption{Deployment from source across 11~microservice applications, with and
  without deployment-time memory. Both configurations use the independent
  health judge; summary-generation cost is included in the with-summary totals.}
  \label{fig:deploy}
\end{figure*}

\subsection{Deployment from Source}
\label{sec:eval:deployment}

This experiment tests whether \sys can bring applications from source to
deployment, and which design decisions
make that process reliable and efficient.

\paragraph{Setup.}
For each application in Table~\ref{tab:eval-apps}, we start from a
fixed source commit and an empty Kubernetes namespace.
Before the experiment, we remove the pre-existing Kubernetes deployment
configs and Docker setup in the source repositories to emulate initial deployment of a new application.
The agent must infer how to containerize
each service, build the images, and connect the services into an
end-to-end deployment.
We evaluate the performance of the resulting deployment with application-specific load
generators against the Kubernetes cluster where the application is deployed.

\paragraph{Metrics.}
We report deployment success rate,
deployment time, and LLM token use.
Deployment time includes authoring service Dockerfiles, building service
images, and standing up the services in Kubernetes.
We also report false success declarations:
cases where the deployer stops and claims success but the external load
generator still fails.

\paragraph{Results.}
Figure~\ref{fig:deploy} reports the deployment time, iteration count, and token usage of \sys with and without operational memory. Both configurations successfully deploy all 11 applications under an iteration budget of 100.
We can see that operational memory reduces deployment time and token use by 47.5\%, while reducing iterations by 2.5$\times$.

The largest gains are from applications that face complex build and deployment steps, such as train-ticket and socialNetwork.
These are applications that use static languages, such as Java and C++, in their microservice implementations,
have high service count, and have a complex dependency graph between services.
Before the Kubernetes deployment can succeed, the agent must first infer the
correct build procedure for each service. Failed builds force it to read
compiler or build-system logs, revise Dockerfiles, and retry.

The architecture summary, in particular, preserves the high-level
deployment plan across these token-intensive repair loops.
Without the summary, agents often spend context rediscovering service
roles, build commands, and dependency edges while investigating build
errors.
These sessions are long enough that the deployer can exhaust the
200k-token context window for Sonnet 4.6, the model used in our evaluation,
resulting in compaction and loss of important context.
With the summary, \sys records how it expects each service to be built
and connected before iterating on low-level failures, so later attempts
resume from the deployment structure rather than from raw logs.

\begin{table}[t]
  \centering
  \footnotesize
  \caption{Health-judge ablation. Final success is measured by the same external load generators.}
  \label{tab:health-judge-ablation}
  \begin{tabular}{lccc}
    \hline
    Configuration & Success & Mean time & Mean tokens \\
    \hline
    Single-agent & 4/11 & 46m & 6.2M \\
    With judge & 11/11 & 107m & 21.9M \\
    \hline
  \end{tabular}
\end{table}

\paragraph{Effect of health judge.}
We separately ablate the independent health judge.
Table~\ref{tab:health-judge-ablation} reports externally validated
deployment success with and without it. An application counts as successful only if the final
deployment passes the load generator.
Time and token values are means across the 11~deployment applications.

Without the health judge (single-agent), the
deployer is too eager to stop: every run eventually declared success, reducing time and token use, yet
only four deployments passed external validation.
The failed runs were partial deployments.
For example, some runs omitted required databases. Some large applications, such as train-ticket, were declared complete before all
services were deployed.
As the single agent spends more context on build and deployment details,
it can lose the high-level health objective and accept local progress as
end-to-end success.
The independent judge prevents this drift by keeping validation anchored
to the user-facing load generator. With the judge, all 11 deployments
passed the same external success criterion.

\subsection{Memory Transfer}
\label{sec:eval:memory-transfer}

We drill down into how \sys's memory applies to new incidents in two scenarios.
In the \emph{overlap} case, a new incident repeats a fault \sys has already seen.
Does its memory resolve the incident faster and cheaper?
In the \emph{no-overlap} case, a new incident shares nothing with prior memory.
Does carrying that memory hurt resolution time, accuracy, or token usage?

\paragraph{Setup.}
We divide incidents generated from SREGym's fault templates into a 50-incident
training set and two 30-incident test sets: the overlap set reuses templates from
the training set but with different parameter bindings, so each incident repeats a
known root cause without being identical, and the no-overlap set shares no
templates.
We build memory by running \sys on the training set, then freeze it during the
test runs so agents cannot modify it, isolating its effect from in-run learning.
We compare against Memoryless on both test sets, using the metrics of
\S\ref{sec:eval:effectiveness}.

\begin{figure}[t]
  \centering
  \begin{subfigure}[t]{\columnwidth}
    \centering
    \includegraphics[width=0.49\linewidth]{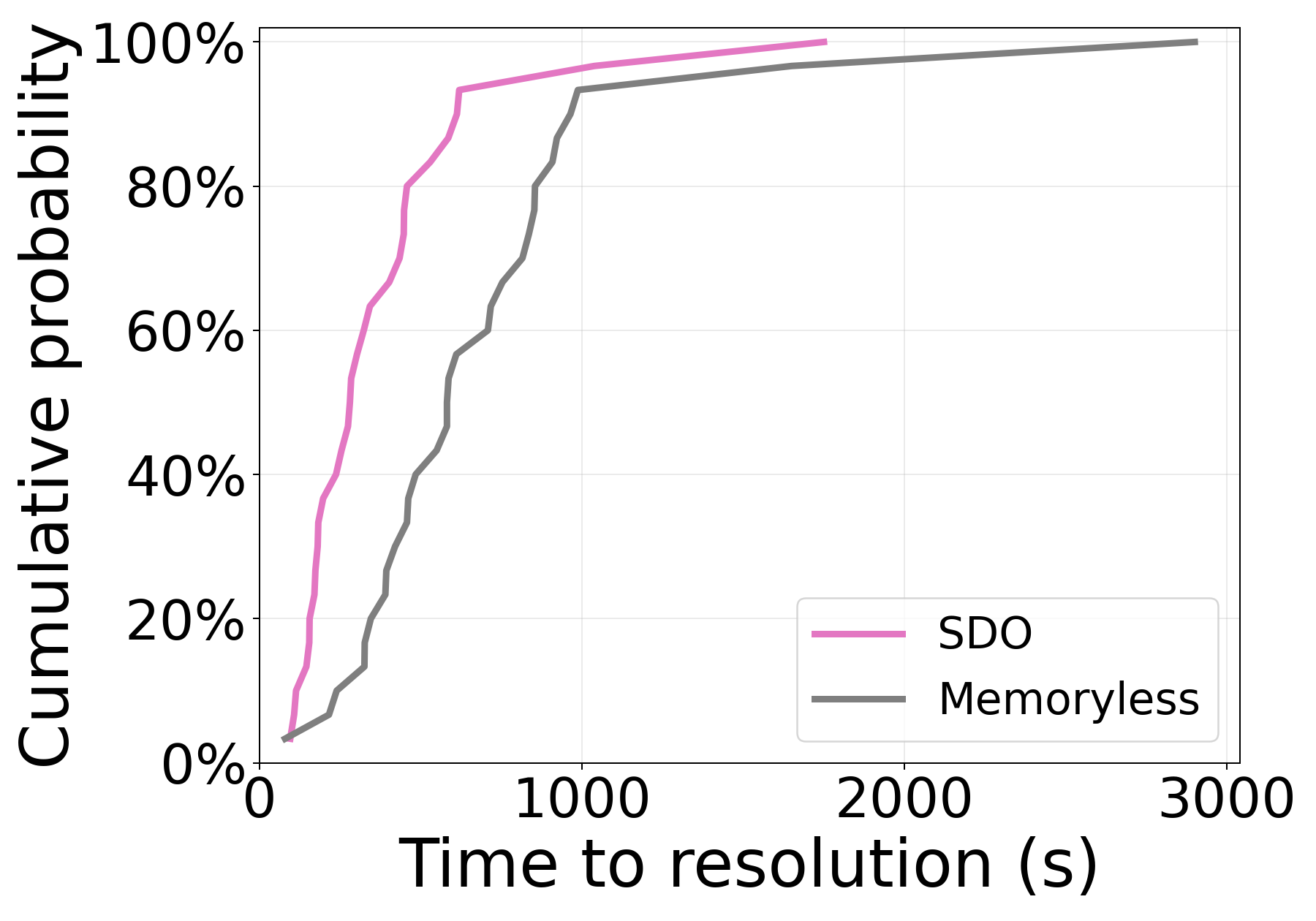}\hfill
    \includegraphics[width=0.49\linewidth]{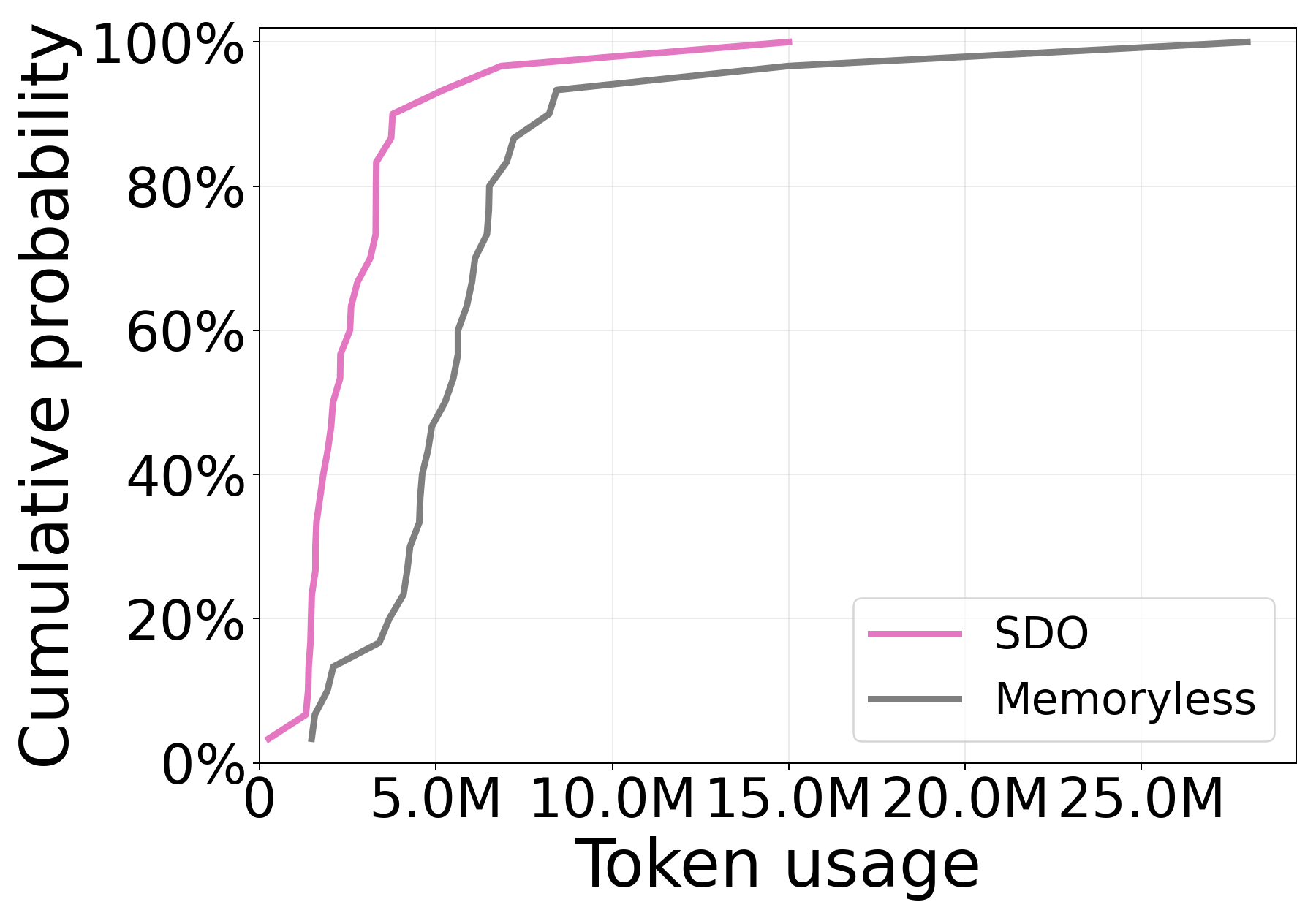}
    \caption{Overlapping root causes.}
    \label{fig:mem-transfer-overlap}
  \end{subfigure}

  \vspace{0.5em}
  \begin{subfigure}[t]{\columnwidth}
    \centering
    \includegraphics[width=0.49\linewidth]{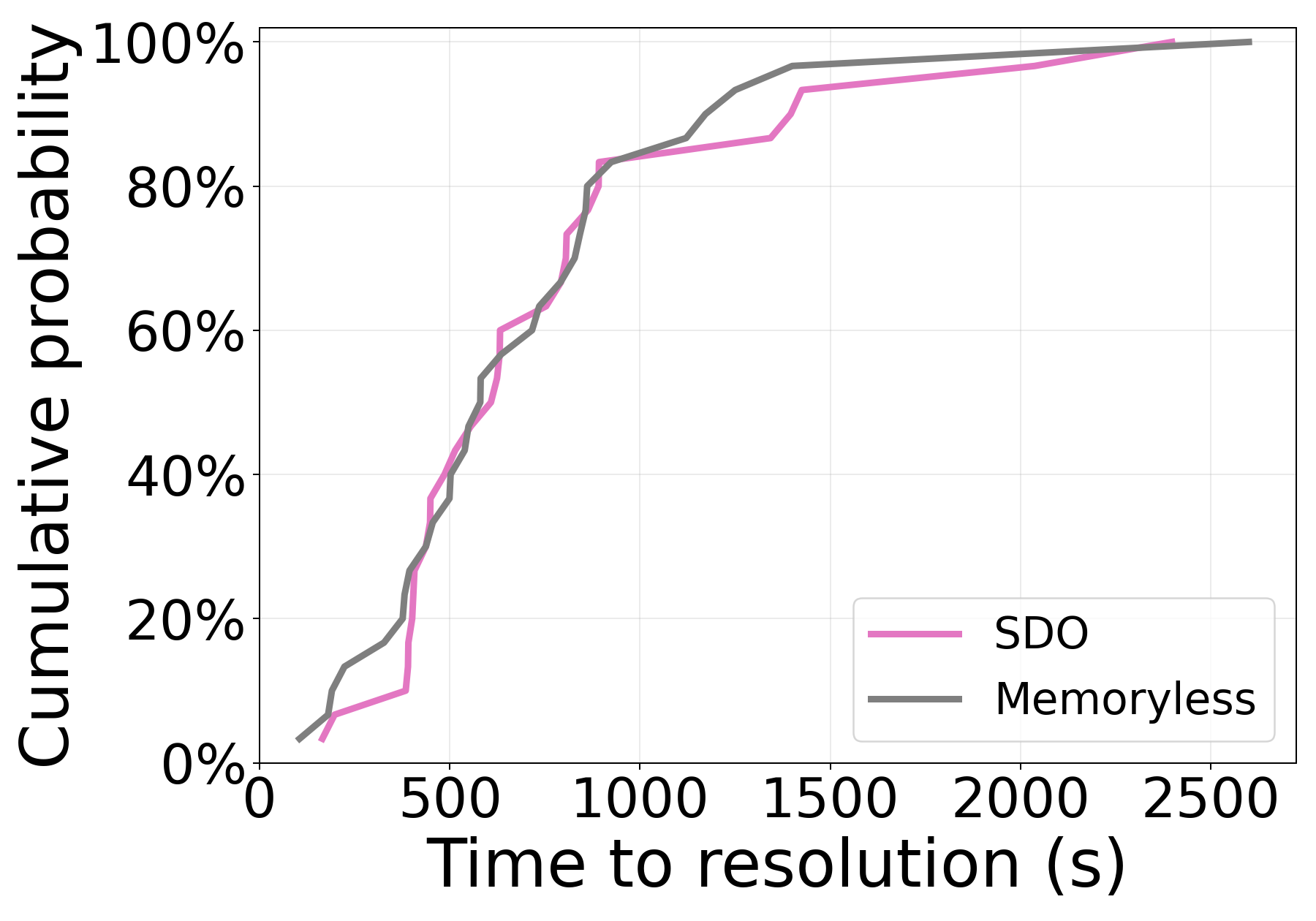}\hfill
    \includegraphics[width=0.49\linewidth]{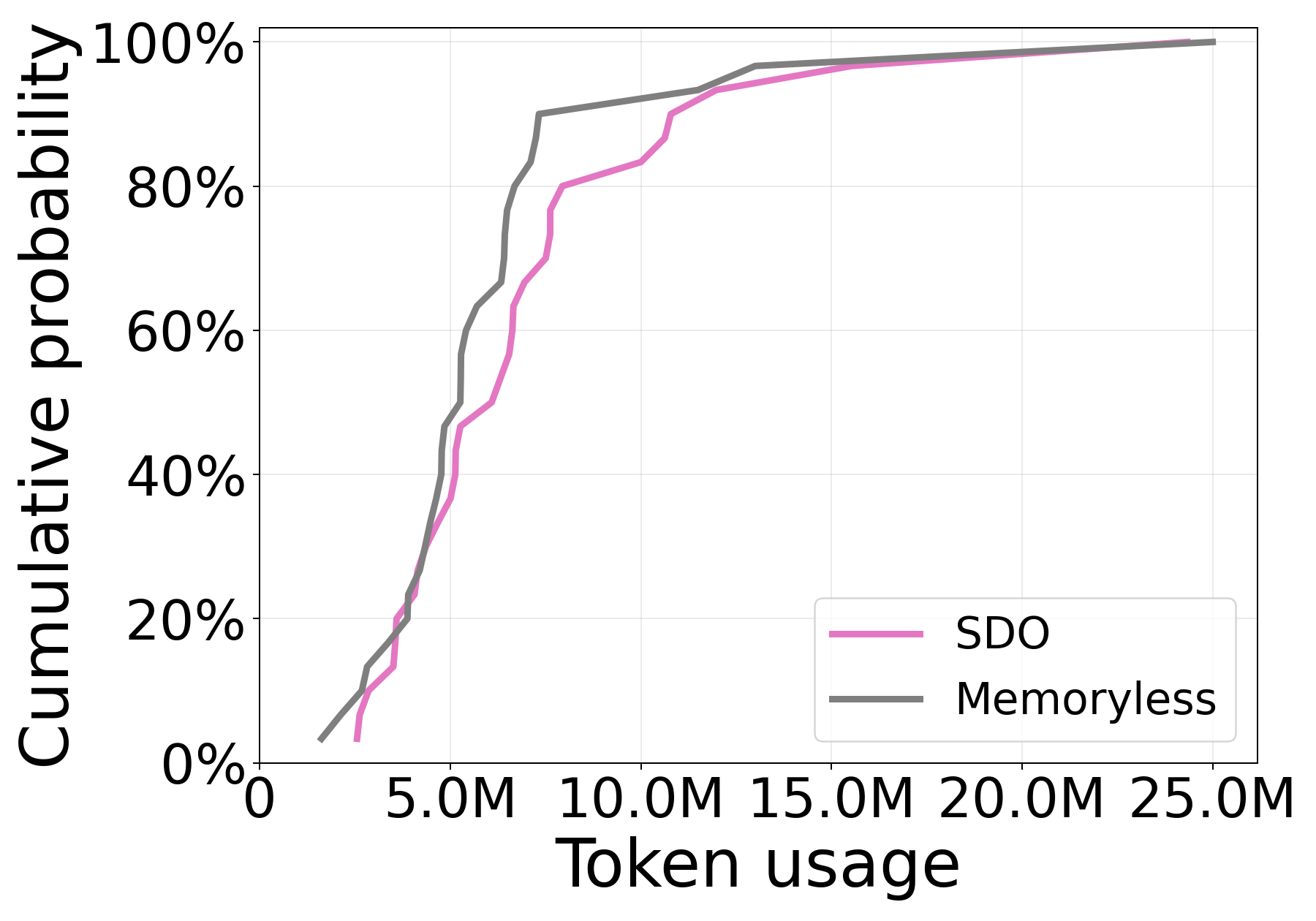}
    \caption{No-overlap root causes.}
    \label{fig:mem-transfer-no-overlap}
  \end{subfigure}
  \caption{CDFs of time to resolution (left) and token usage (right) on the
  transfer test sets.}
  \label{fig:memory-transfer}
\end{figure}

\paragraph{Results.}
On recurring faults, \sys resolves incidents faster and cheaper than Memoryless.
In Figure~\ref{fig:mem-transfer-overlap}, \sys's time-to-resolution and
token-usage CDFs lie to the left of Memoryless at every percentile: the median
incident resolves in 282\,s rather than 584\,s and consumes 2.2M rather than
5.4M tokens, resolving 2.1$\times$ faster and using 2.5$\times$ fewer tokens.
The tail in \sys's CDFs reflects cases where the agent falls back to open-ended diagnosis and performs similarly to Memoryless; 4 of the 30 incidents fall on this tail.

In contrast, the Memoryless agent spends most time and tokens investigating incidents. 
Solve rates are similar across Memoryless and \sys agents.
A firing detector accelerates resolution but does not guarantee end-to-end
success: because each incident composes multiple faults, both Memoryless and
\sys agents sometimes diagnose and mitigate only a subset of an incident's
faults, leaving incidents not fully resolved even when a detector surfaced a relevant playbook.

On faults with no overlap to prior memory, \sys maintains accuracy and
time-to-resolution similar to those of the Memoryless agents (Figure~\ref{fig:mem-transfer-no-overlap}).
In 9 of 30 novel incidents, detectors surface playbooks.
However, because incidents in this set share no root cause with the training set, every
surfaced playbook is a false positive.
The agent inspects and discards the false positives without being misled.
Accuracy and resolution time are similar to Memoryless (Table~\ref{tab:memory-transfer-solve-rates}), so a spurious playbook
adds some inspection but does not yield a wrong diagnosis or a slower fix.
Because we freeze memory in the test set, \sys agents cannot prune mis-firing
detectors as they would in live operation (\S\ref{sec:eval:effectiveness}).

\sys sees a modest token overhead compared to Memoryless: a
17.7\% increase at the median.
However, this did not translate to increased time-to-resolution.
First, an \sys agent dismissing a suggested diagnosis is not fully wasted work:
the agent still learned information about the cluster that is useful for
subsequent diagnostic steps within the same incident. Second,
we find that the \sys agent writes more analysis scripts than the Memoryless one---in 83\% of incidents versus 53\% for the
Memoryless agent. Scripts can reduce execution time. Rather than issue one
command per tool call and wait for its response, the agent batches the requisite
commands into a single script and retrieves only the output it needs. 
Two factors contribute to \sys agents' propensity to write scripts: (1) \sys agents contain instructions on writing scripts as part of their memory instructions; (2) they see example scripts from the playbooks in frozen operational memory.

\begin{table}[t]
  \centering
  \footnotesize
  \caption{Diagnosis, mitigation, and end-to-end solve rates on the
  transfer test sets.}
  \label{tab:memory-transfer-solve-rates}
  \begin{tabular}{llccc}
    \hline
    Test set & Configuration & Diagnosis & Mitigation & End-to-end \\
    \hline
    Overlap    & Memoryless & 70\% & 73.3\% & 63.3\% \\
               & \sys       & 66.7\% & 73.3\% & 60\% \\
    \hline
    No-overlap & Memoryless & 63.3\% & 76.7\% & 56.7\% \\
               & \sys       & 60\% & 73.3\% & 56.7\% \\
    \hline
  \end{tabular}
\end{table}

\subsection{Retrieval Accuracy and Cost}
\label{sec:eval:retrieval}

We investigate whether detectors retrieve the right playbooks.
Unlike \S\ref{sec:eval:effectiveness} and \S\ref{sec:eval:memory-transfer},
which measure end-to-end outcomes, this experiment isolates retrieval: all
configurations retrieve from the same playbook corpus learned by \sys, so any
difference in accuracy reflects the mechanism rather than the stored memory.

\paragraph{Setup.}
We reuse the frozen memory and the overlap and no-overlap test sets of
\S\ref{sec:eval:memory-transfer}.
On the \emph{overlap} set, a relevant playbook exists in the corpus, so we
report precision and recall over pooled matching decisions. Since an incident
can match several playbooks, we pool the true positives, false positives, and
false negatives across all incidents. We then compute a single precision and
recall from the aggregate counts. On the
\emph{no-overlap} set, no playbook matches the
incident, so we report the false-positive rate: the fraction of incidents for
which a mechanism surfaces any playbook.

Holding the playbooks fixed, we compare three retrieval mechanisms.
\emph{Playbook-Embed} and \emph{Playbook-LLM} apply the retrieval mechanisms
of History-Embed and History-LLM (\S\ref{sec:eval:effectiveness}) to \sys's
playbooks instead of raw incident histories, and \emph{Detectors} uses the
agent-authored detector code \sys stores alongside the playbooks.
Because the benchmark defines which faults each incident contains, we label the
playbooks relevant to each incident as ground truth, then run each mechanism on
the retrieval step alone and score what it surfaces against these labels.
We score a single retrieval at incident onset; a live end-to-end run instead
retrieves repeatedly as the agent tests hypotheses, and \sys's subsequent
retrievals take an LLM-based slow path equivalent to Playbook-LLM.

\begin{table}[t]
  \centering
  \footnotesize
  \caption{Retrieval accuracy and cost across mechanisms. Time and Tokens are median.
  $\uparrow$/$\downarrow$: higher/lower is better.}
  \label{tab:retrieval}
  \begin{tabular}{l|cc|c|cc}
    \hline
    \multirow{3}{*}{Mechanism} & \multicolumn{3}{|c}{Accuracy} & \multicolumn{2}{|c}{Cost} \\
    \cline{2-4} \cline{5-6}
     & \multicolumn{2}{|c}{Overlap} & \multicolumn{1}{|c}{No-overlap} & \multicolumn{2}{|c}{} \\
    \cline{2-3} \cline{4-4}
     & Prec.\,$\uparrow$ & Rec.\,$\uparrow$ & FP\,$\downarrow$ & Time\,$\downarrow$ & Tok.\,$\downarrow$ \\
    \hline
    Playbook-Embed & 0.08 & 0.40 & 1.00 & 64.4\,s & 62k \\
    Playbook-LLM   & 0.53 & 0.77 & 0.17 & 483.4\,s & 148k \\
    \sys Detectors & 0.48 & 0.73 & 0.30 & 10.8\,s & 0 \\
    \hline
  \end{tabular}
\end{table}

\paragraph{Results.}
Table~\ref{tab:retrieval} reports the accuracy and cost of the three mechanisms.
Detectors retrieve about as accurately as Playbook-LLM and dominate
Playbook-Embed on every metric, while incurring no LLM cost at retrieval time.
Embedding is the worst of both worlds: it pays LLM tokens yet retrieves least
accurately. Relative to Playbook-LLM, detectors give up some precision in exchange for roughly 45$\times$ less time and no tokens.
Precision is modest in absolute terms across all mechanisms because an incident
can match several playbooks and, more importantly, the same symptom often maps
to many candidate playbooks with distinct root causes. Retrieval cannot separate
these candidates on symptoms alone; discriminating among them is the diagnostic
work \sys leaves to the responder, which inspects the live cluster and rejects the
playbooks that do not apply (\S\ref{sec:eval:memory-transfer}).

The mechanisms differ in how they find relevant playbooks, which drives both 
accuracy and cost gaps.
Playbook-Embed relies on text similarity, but incidents describe concrete
symptoms while playbooks are written generically, so the two share little text
and it misses most relevant playbooks. It nonetheless uses LLM tokens:
the agent must still explore the cluster
to form a query, which costs both time and tokens.
Playbook-LLM is more accurate because the agent performs retrieval itself,
inspecting live cluster state and correlating it with playbook content by
in-context reasoning rather than text overlap.
This achieves better accuracy than Detectors at the expense of retrieval time
and token cost.
Detectors instead encode that correlation in advance: a prior agent writes
detector code alongside the playbook, offloading the reasoning into a
deterministic predicate over cluster state that hands the next agent a relevant
playbook and a starting point to investigate from. Because the predicate runs
without an LLM, retrieval costs no tokens; \sys runs each detector twice to
exclude transient failures, adding only a 10\,s delay across two detector runs.

On the no-overlap set, where no playbook is relevant, embedding surfaces
playbooks on every incident because it always returns its top five by
similarity, without reasoning about whether they apply. Playbook-LLM avoids
spurious retrieval best, followed by detectors, which lack the LLM reasoning to
rule out playbooks with similar symptoms and so report a higher false positive
rate. \S\ref{sec:eval:memory-transfer} shows that the \sys agent dismisses these
irrelevant playbooks effectively.

\section{Related Work}
\label{sec:related}

\paragraph{LLM agents for system operations.}
A growing line of work applies LLM-based agents to system tasks.
Stratus~\cite{stratus} diagnoses faults on microservice deployments, with a
focus on operational safety; RCACopilot~\cite{rcacopilot}
predicts a root-cause category for an alert; and FLASH~\cite{flash}
follows human-authored troubleshooting guides to diagnose recurring
incidents.
Benchmarks such as AIOpsLab~\cite{aiopslab} and ITBench~\cite{itbench}
evaluate such agents on already-deployed, faulted clusters.
Evolutionary systems such as AlphaEvolve~\cite{alphaevolve},
ADRS~\cite{adrs}, and Glia~\cite{glia} instead optimize components by
iterating over a population of candidate solutions.
These systems either keep state within a single session or consume
human-authored knowledge, and the evolutionary ones retain a population
but discard negative lessons.
\sys instead persists each session as memory that is auditable, directly
invocable as agent-authored scripts, and retrieved by checking live
cluster state rather than by symptom similarity, so it stays aligned with
a continuously changing deployment.

\paragraph{Self-improving LLM agents.}
A separate line of work uses environment feedback to improve agents over
time.
Voyager~\cite{voyager} grows an executable skill library;
Reflexion~\cite{reflexion} reflects verbally across repeated attempts at
the same task; ExpeL~\cite{expel} and Agent Workflow Memory~\cite{awm}
distill cross-task lessons as natural-language text; and RLAIF and
related techniques fine-tune the model on outcome-derived preferences.
Recursive self-improvement systems go further, rewriting their own code
or policy to become more capable agents in general: STOP~\cite{stop},
G{\"o}del Agent~\cite{godelagent}, and the Darwin G{\"o}del
Machine~\cite{dgm} edit their own implementation, in the lineage of
Schmidhuber's G{\"o}del machine~\cite{godelmachine}.
\sys accumulates executable artifacts like Voyager rather than the text
lessons of Reflexion, ExpeL, and AWM, but its self-definition is
deliberately bounded: it holds the model and a fixed detector API
constant and revises only a deployment's operational memory, so unlike
the self-rewriting systems every change stays auditable and reversible.
Unlike all of this work, \sys operates in production and treats the
running cluster as its supervisory signal.

\paragraph{Memory and long context for LLMs.}
General-purpose memory systems~\cite{memgpt} and retrieval-augmented
generation~\cite{rag} externalize knowledge into searchable stores, but
optimize for dialogue and question-answering, where the units of memory
are short facts retrieved by embedding similarity; serving-layer caches
reuse past responses the same way to cut cost~\cite{iccache}.
\sys's knowledge base differs structurally: a per-application world model
grounds every session, and entries are parameterized causal chains with
executable scripts, read on demand by a single role, rather than text
documents.
The parameterized form keeps literal entity names out of the diagnostic
context, which is what makes \sys robust to deployment drift where
generic RAG over post-mortems is not.
Larger context windows do not subsume this: loading raw history into
every session is costly, dilutes attention, and still lacks the
parameterization needed to survive drift.

\paragraph{Runbooks and incident management.}
\sys's playbooks build on practitioner-authored runbooks and standard
operating procedures~\cite{sre,sreworkbook}, but are authored and refined
by the agent itself, parameterized over the deployment's entities, and
coupled with executable scripts.
Prior systems recommend an existing human-authored guide for an
incident~\cite{tsgrec,flash}; \sys instead authors the guidance, while
remaining able to import human runbooks as seed material.

\section{Conclusion}

\sys shows that autonomous operators can improve with experience by turning incident resolutions into operational memory. Across 11 microservice applications, compared to a memoryless agent, \sys achieves the highest end-to-end incident solve rate, reduces application deployment time and token use by up to 47.5\%, and resolves recurring faults up to 2.1$\times$ faster, while using 2.5$\times$ fewer tokens. These results suggest that the right memory for SRE agents is not raw incident history, but operating procedure: detectors that retrieve applicable knowledge, playbooks that encode reusable repairs, and architecture summaries that keep both grounded in the evolving application.

{\footnotesize \bibliographystyle{ACM-Reference-Format}
  \bibliography{paper}}


\begin{thebibliography}{67}


\ifx \showCODEN    \undefined \def \showCODEN     #1{\unskip}     \fi
\ifx \showDOI      \undefined \def \showDOI       #1{#1}\fi
\ifx \showISBNx    \undefined \def \showISBNx     #1{\unskip}     \fi
\ifx \showISBNxiii \undefined \def \showISBNxiii  #1{\unskip}     \fi
\ifx \showISSN     \undefined \def \showISSN      #1{\unskip}     \fi
\ifx \showLCCN     \undefined \def \showLCCN      #1{\unskip}     \fi
\ifx \shownote     \undefined \def \shownote      #1{#1}          \fi
\ifx \showarticletitle \undefined \def \showarticletitle #1{#1}   \fi
\ifx \showURL      \undefined \def \showURL       {\relax}        \fi
\providecommand\bibfield[2]{#2}
\providecommand\bibinfo[2]{#2}
\providecommand\natexlab[1]{#1}
\providecommand\showeprint[2][]{arXiv:#2}

\bibitem[age(2025)]%
        {agentskills}
 \bibinfo{year}{2025}\natexlab{}.
\newblock \bibinfo{title}{Agent Skills}.
\newblock \bibinfo{howpublished}{\url{https://agentskills.io}}.
\newblock


\bibitem[aws(2025)]%
        {awsdevopsagent}
 \bibinfo{year}{2025}\natexlab{}.
\newblock \bibinfo{title}{Amazon {DevOps} Agent}.
\newblock \bibinfo{howpublished}{\url{https://aws.amazon.com/devops-agent/}}.
\newblock


\bibitem[azu(2025)]%
        {azuresreagent}
 \bibinfo{year}{2025}\natexlab{}.
\newblock \bibinfo{title}{Azure {SRE} Agent}.
\newblock
  \bibinfo{howpublished}{\url{https://azure.microsoft.com/en-us/products/sre-agent}}.
\newblock


\bibitem[pyd(2025)]%
        {pydanticai}
 \bibinfo{year}{2025}\natexlab{}.
\newblock \bibinfo{title}{{PydanticAI}}.
\newblock \bibinfo{howpublished}{\url{https://ai.pydantic.dev}}.
\newblock


\bibitem[{Amazon Web Services}(2025)]%
        {awsincidentmanager}
\bibfield{author}{\bibinfo{person}{{Amazon Web Services}}.}
  \bibinfo{year}{2025}\natexlab{}.
\newblock \bibinfo{title}{{AWS Systems Manager Incident Manager}}.
\newblock
  \bibinfo{howpublished}{\url{https://aws.amazon.com/systems-manager/features/incident-manager/}}.
\newblock


\bibitem[{Anthropic}(2025)]%
        {mcp}
\bibfield{author}{\bibinfo{person}{{Anthropic}}.}
  \bibinfo{year}{2025}\natexlab{}.
\newblock \bibinfo{title}{{Model Context Protocol Specification}}.
\newblock
  \bibinfo{howpublished}{\url{https://modelcontextprotocol.io/specification/2025-11-25}}.
\newblock


\bibitem[Beyer et~al\mbox{.}(2016)]%
        {sre}
\bibfield{author}{\bibinfo{person}{Betsy Beyer}, \bibinfo{person}{Chris Jones},
  \bibinfo{person}{Jennifer Petoff}, {and} \bibinfo{person}{Niall~Richard
  Murphy}.} \bibinfo{year}{2016}\natexlab{}.
\newblock \bibinfo{booktitle}{\emph{Site Reliability Engineering: How {Google}
  Runs Production Systems}}.
\newblock \bibinfo{publisher}{O'Reilly Media}.
\newblock


\bibitem[Beyer et~al\mbox{.}(2018)]%
        {sreworkbook}
\bibfield{author}{\bibinfo{person}{Betsy Beyer}, \bibinfo{person}{Niall~Richard
  Murphy}, \bibinfo{person}{David~K. Rensin}, \bibinfo{person}{Kent Kawahara},
  {and} \bibinfo{person}{Stephen Thorne}.} \bibinfo{year}{2018}\natexlab{}.
\newblock \bibinfo{booktitle}{\emph{The Site Reliability Workbook: Practical
  Ways to Implement {SRE}}}.
\newblock \bibinfo{publisher}{O'Reilly Media}.
\newblock


\bibitem[Chen et~al\mbox{.}(2026)]%
        {stratus}
\bibfield{author}{\bibinfo{person}{Yinfang Chen}, \bibinfo{person}{Jiaqi Pan},
  \bibinfo{person}{Jackson Clark}, \bibinfo{person}{Yiming Su},
  \bibinfo{person}{Noah Zheutlin}, \bibinfo{person}{Bhavya Bhavya},
  \bibinfo{person}{Rohan Arora}, \bibinfo{person}{Yu Deng},
  \bibinfo{person}{Saurabh Jha}, {and} \bibinfo{person}{Tianyin Xu}.}
  \bibinfo{year}{2026}\natexlab{}.
\newblock \bibinfo{title}{STRATUS: A Multi-agent System for Autonomous
  Reliability Engineering of Modern Clouds}.
\newblock
\newblock
\showeprint[arxiv]{2506.02009}~[cs.DC]
\urldef\tempurl%
\url{https://arxiv.org/abs/2506.02009}
\showURL{%
\tempurl}


\bibitem[Chen et~al\mbox{.}(2025)]%
        {aiopslab}
\bibfield{author}{\bibinfo{person}{Yinfang Chen}, \bibinfo{person}{Manish
  Shetty}, \bibinfo{person}{Gagan Somashekar}, \bibinfo{person}{Minghua Ma},
  \bibinfo{person}{Yogesh Simmhan}, \bibinfo{person}{Jonathan Mace},
  \bibinfo{person}{Chetan Bansal}, \bibinfo{person}{Rujia Wang}, {and}
  \bibinfo{person}{Saravan Rajmohan}.} \bibinfo{year}{2025}\natexlab{}.
\newblock \bibinfo{title}{{AIOpsLab}: A Holistic Framework to Evaluate AI
  Agents for Enabling Autonomous Clouds}.
\newblock
\newblock
\showeprint[arxiv]{2501.06706}~[cs.SE]
\urldef\tempurl%
\url{https://arxiv.org/abs/2501.06706}
\showURL{%
\tempurl}


\bibitem[Chen et~al\mbox{.}(2024)]%
        {rcacopilot}
\bibfield{author}{\bibinfo{person}{Yinfang Chen}, \bibinfo{person}{Huaibing
  Xie}, \bibinfo{person}{Minghua Ma}, \bibinfo{person}{Yu Kang},
  \bibinfo{person}{Xin Gao}, \bibinfo{person}{Liu Shi}, \bibinfo{person}{Yunjie
  Cao}, \bibinfo{person}{Xuedong Gao}, \bibinfo{person}{Hao Fan},
  \bibinfo{person}{Ming Wen}, \bibinfo{person}{Jun Zeng},
  \bibinfo{person}{Supriyo Ghosh}, \bibinfo{person}{Xuchao Zhang},
  \bibinfo{person}{Chaoyun Zhang}, \bibinfo{person}{Qingwei Lin},
  \bibinfo{person}{Saravan Rajmohan}, \bibinfo{person}{Dongmei Zhang}, {and}
  \bibinfo{person}{Tianyin Xu}.} \bibinfo{year}{2024}\natexlab{}.
\newblock \showarticletitle{Automatic Root Cause Analysis via Large Language
  Models for Cloud Incidents}. In \bibinfo{booktitle}{\emph{European Conference
  on Computer Systems (EuroSys)}}.
\newblock


\bibitem[Chen et~al\mbox{.}(2020)]%
        {linkedincidents}
\bibfield{author}{\bibinfo{person}{Yujun Chen}, \bibinfo{person}{Xian Yang},
  \bibinfo{person}{Hang Dong}, \bibinfo{person}{Xiaoting He},
  \bibinfo{person}{Hongyu Zhang}, \bibinfo{person}{Qingwei Lin},
  \bibinfo{person}{Junjie Chen}, \bibinfo{person}{Pu Zhao}, \bibinfo{person}{Yu
  Kang}, \bibinfo{person}{Feng Gao}, \bibinfo{person}{Zhangwei Xu}, {and}
  \bibinfo{person}{Dongmei Zhang}.} \bibinfo{year}{2020}\natexlab{}.
\newblock \showarticletitle{Identifying Linked Incidents in Large-Scale Online
  Service Systems}. In \bibinfo{booktitle}{\emph{ACM Joint European Software
  Engineering Conference and Symposium on the Foundations of Software
  Engineering (ESEC/FSE)}}.
\newblock


\bibitem[Cheng et~al\mbox{.}(2025)]%
        {adrs}
\bibfield{author}{\bibinfo{person}{Audrey Cheng}, \bibinfo{person}{Shu Liu},
  \bibinfo{person}{Melissa Pan}, \bibinfo{person}{Zhifei Li},
  \bibinfo{person}{Shubham Agarwal}, \bibinfo{person}{Mert Cemri},
  \bibinfo{person}{Bowen Wang}, \bibinfo{person}{Alexander Krentsel},
  \bibinfo{person}{Tian Xia}, \bibinfo{person}{Jongseok Park},
  \bibinfo{person}{Shuo Yang}, \bibinfo{person}{Jeff Chen},
  \bibinfo{person}{Lakshya Agrawal}, \bibinfo{person}{Ashwin Naren},
  \bibinfo{person}{Shulu Li}, \bibinfo{person}{Ruiying Ma},
  \bibinfo{person}{Aditya Desai}, \bibinfo{person}{Jiarong Xing},
  \bibinfo{person}{Koushik Sen}, \bibinfo{person}{Matei Zaharia}, {and}
  \bibinfo{person}{Ion Stoica}.} \bibinfo{year}{2025}\natexlab{}.
\newblock \bibinfo{title}{Let the Barbarians In: How AI Can Accelerate Systems
  Performance Research}.
\newblock
\newblock
\showeprint[arxiv]{2512.14806}~[cs.SE]
\urldef\tempurl%
\url{https://arxiv.org/abs/2512.14806}
\showURL{%
\tempurl}


\bibitem[Clark et~al\mbox{.}(2026)]%
        {sregym}
\bibfield{author}{\bibinfo{person}{Jackson Clark}, \bibinfo{person}{Yiming Su},
  \bibinfo{person}{Saad Mohammad~Rafid Pial}, \bibinfo{person}{Yifang Tian},
  \bibinfo{person}{Lily Gniedziejko}, \bibinfo{person}{Hans-Arno Jacobsen},
  \bibinfo{person}{Yinfang Chen}, {and} \bibinfo{person}{Tianyin Xu}.}
  \bibinfo{year}{2026}\natexlab{}.
\newblock \bibinfo{title}{SREGym: A Live Benchmark for AI SRE Agents with
  High-Fidelity Failure Scenarios}.
\newblock
\newblock
\showeprint[arxiv]{2605.07161}~[cs.AI]
\urldef\tempurl%
\url{https://arxiv.org/abs/2605.07161}
\showURL{%
\tempurl}


\bibitem[Cotroneo et~al\mbox{.}(2019)]%
        {openstackfaults}
\bibfield{author}{\bibinfo{person}{Domenico Cotroneo}, \bibinfo{person}{Luigi
  De~Simone}, \bibinfo{person}{Pietro Liguori}, \bibinfo{person}{Roberto
  Natella}, {and} \bibinfo{person}{Nematollah Bidokhti}.}
  \bibinfo{year}{2019}\natexlab{}.
\newblock \showarticletitle{How Bad Can a Bug Get? An Empirical Analysis of
  Software Failures in the {OpenStack} Cloud Computing Platform}. In
  \bibinfo{booktitle}{\emph{ACM Joint European Software Engineering Conference
  and Symposium on the Foundations of Software Engineering (ESEC/FSE)}}.
\newblock


\bibitem[{Datadog}(2025)]%
        {datadog}
\bibfield{author}{\bibinfo{person}{{Datadog}}.}
  \bibinfo{year}{2025}\natexlab{}.
\newblock \bibinfo{title}{{Datadog: Cloud Monitoring as a Service}}.
\newblock \bibinfo{howpublished}{\url{https://www.datadoghq.com}}.
\newblock


\bibitem[Dogga et~al\mbox{.}(2023)]%
        {autoarts}
\bibfield{author}{\bibinfo{person}{Pradeep Dogga}, \bibinfo{person}{Chetan
  Bansal}, \bibinfo{person}{Richard Costleigh}, \bibinfo{person}{Gopinath
  Jayagopal}, \bibinfo{person}{Suman Nath}, {and} \bibinfo{person}{Xuchao
  Zhang}.} \bibinfo{year}{2023}\natexlab{}.
\newblock \showarticletitle{{AutoARTS}: Taxonomy, Insights and Tools for Root
  Cause Labelling of Incidents in {Microsoft Azure}}. In
  \bibinfo{booktitle}{\emph{USENIX Annual Technical Conference (ATC)}}.
\newblock


\bibitem[{Dynatrace}(2025)]%
        {dynatrace}
\bibfield{author}{\bibinfo{person}{{Dynatrace}}.}
  \bibinfo{year}{2025}\natexlab{}.
\newblock \bibinfo{title}{{Dynatrace: AI-Powered Observability Platform}}.
\newblock \bibinfo{howpublished}{\url{https://www.dynatrace.com}}.
\newblock


\bibitem[{FireHydrant}(2025)]%
        {firehydrant}
\bibfield{author}{\bibinfo{person}{{FireHydrant}}.}
  \bibinfo{year}{2025}\natexlab{}.
\newblock \bibinfo{title}{{FireHydrant: Incident Management and Response
  Platform}}.
\newblock \bibinfo{howpublished}{\url{https://firehydrant.com}}.
\newblock


\bibitem[{FleetCast}(2025)]%
        {fleetcast}
\bibfield{author}{\bibinfo{person}{{FleetCast}}.}
  \bibinfo{year}{2025}\natexlab{}.
\newblock \bibinfo{title}{{FleetCast}}.
\newblock \bibinfo{howpublished}{\url{https://github.com/lilygn/FleetCast}}.
\newblock


\bibitem[Gan et~al\mbox{.}(2019)]%
        {deathstarbench}
\bibfield{author}{\bibinfo{person}{Yu Gan}, \bibinfo{person}{Yanqi Zhang},
  \bibinfo{person}{Dailun Cheng}, \bibinfo{person}{Ankitha Shetty},
  \bibinfo{person}{Priyal Rathi}, \bibinfo{person}{Nayan Katarki},
  \bibinfo{person}{Ariana Bruno}, \bibinfo{person}{Justin Hu},
  \bibinfo{person}{Brian Ritchken}, \bibinfo{person}{Brendon Jackson},
  \bibinfo{person}{Kelvin Hu}, \bibinfo{person}{Meghna Pancholi},
  \bibinfo{person}{Yuan He}, \bibinfo{person}{Brett Clancy},
  \bibinfo{person}{Chris Colen}, \bibinfo{person}{Fukang Wen},
  \bibinfo{person}{Catherine Leung}, \bibinfo{person}{Siyuan Wang},
  \bibinfo{person}{Leon Zaruvinsky}, \bibinfo{person}{Mateo Espinosa},
  \bibinfo{person}{Rick Lin}, \bibinfo{person}{Zhongling Liu},
  \bibinfo{person}{Jake Padilla}, {and} \bibinfo{person}{Christina
  Delimitrou}.} \bibinfo{year}{2019}\natexlab{}.
\newblock \showarticletitle{An Open-Source Benchmark Suite for Microservices
  and Their Hardware-Software Implications for Cloud \& Edge Systems}. In
  \bibinfo{booktitle}{\emph{Proceedings of the Twenty-Fourth International
  Conference on Architectural Support for Programming Languages and Operating
  Systems}} (Providence, RI, USA) \emph{(\bibinfo{series}{ASPLOS '19})}.
  \bibinfo{publisher}{Association for Computing Machinery},
  \bibinfo{address}{New York, NY, USA}, \bibinfo{pages}{3–18}.
\newblock
\showISBNx{9781450362405}
\urldef\tempurl%
\url{https://doi.org/10.1145/3297858.3304013}
\showDOI{\tempurl}


\bibitem[Ghosh et~al\mbox{.}(2022)]%
        {fightincidents}
\bibfield{author}{\bibinfo{person}{Supriyo Ghosh}, \bibinfo{person}{Manish
  Shetty}, \bibinfo{person}{Chetan Bansal}, {and} \bibinfo{person}{Suman
  Nath}.} \bibinfo{year}{2022}\natexlab{}.
\newblock \showarticletitle{How to Fight Production Incidents? An Empirical
  Study on a Large-scale Cloud Service}. In \bibinfo{booktitle}{\emph{ACM
  Symposium on Cloud Computing (SoCC)}}.
\newblock


\bibitem[{Google Cloud}(2025)]%
        {gcloudops}
\bibfield{author}{\bibinfo{person}{{Google Cloud}}.}
  \bibinfo{year}{2025}\natexlab{}.
\newblock \bibinfo{title}{{Introduction to Google Cloud's Operations Suite}}.
\newblock
  \bibinfo{howpublished}{\url{https://cloud.google.com/blog/topics/developers-practitioners/introduction-google-clouds-operations-suite}}.
\newblock


\bibitem[{Google Cloud Platform}(2025)]%
        {onlineboutique}
\bibfield{author}{\bibinfo{person}{{Google Cloud Platform}}.}
  \bibinfo{year}{2025}\natexlab{}.
\newblock \bibinfo{title}{Online Boutique}.
\newblock
  \bibinfo{howpublished}{\url{https://github.com/GoogleCloudPlatform/microservices-demo}}.
\newblock


\bibitem[{Grafana Labs}(2025)]%
        {grafana}
\bibfield{author}{\bibinfo{person}{{Grafana Labs}}.}
  \bibinfo{year}{2025}\natexlab{}.
\newblock \bibinfo{title}{{Grafana: The Open Observability Platform}}.
\newblock \bibinfo{howpublished}{\url{https://grafana.com}}.
\newblock


\bibitem[Gunawi et~al\mbox{.}(2014)]%
        {cloudbugs}
\bibfield{author}{\bibinfo{person}{Haryadi~S. Gunawi}, \bibinfo{person}{Mingzhe
  Hao}, \bibinfo{person}{Tanakorn Leesatapornwongsa}, \bibinfo{person}{Tiratat
  Patana-anake}, \bibinfo{person}{Thanh Do}, \bibinfo{person}{Jeffry
  Adityatama}, \bibinfo{person}{Kurnia~J. Eliazar}, \bibinfo{person}{Agung
  Laksono}, \bibinfo{person}{Jeffrey~F. Lukman}, \bibinfo{person}{Vincentius
  Martin}, {and} \bibinfo{person}{Anang~D. Satria}.}
  \bibinfo{year}{2014}\natexlab{}.
\newblock \showarticletitle{What Bugs Live in the Cloud? A Study of 3000+
  Issues in Cloud Systems}. In \bibinfo{booktitle}{\emph{ACM Symposium on Cloud
  Computing (SoCC)}}.
\newblock


\bibitem[Gunawi et~al\mbox{.}(2016)]%
        {cloudstop}
\bibfield{author}{\bibinfo{person}{Haryadi~S. Gunawi}, \bibinfo{person}{Mingzhe
  Hao}, \bibinfo{person}{Riza~O. Suminto}, \bibinfo{person}{Agung Laksono},
  \bibinfo{person}{Anang~D. Satria}, \bibinfo{person}{Jeffry Adityatama}, {and}
  \bibinfo{person}{Kurnia~J. Eliazar}.} \bibinfo{year}{2016}\natexlab{}.
\newblock \showarticletitle{Why Does the Cloud Stop Computing? Lessons from
  Hundreds of Service Outages}. In \bibinfo{booktitle}{\emph{ACM Symposium on
  Cloud Computing (SoCC)}}.
\newblock


\bibitem[Hamadanian et~al\mbox{.}(2025)]%
        {glia}
\bibfield{author}{\bibinfo{person}{Pouya Hamadanian}, \bibinfo{person}{Pantea
  Karimi}, \bibinfo{person}{Arash Nasr-Esfahany}, \bibinfo{person}{Kimia
  Noorbakhsh}, \bibinfo{person}{Joseph Chandler}, \bibinfo{person}{Ali
  ParandehGheibi}, \bibinfo{person}{Mohammad Alizadeh}, {and}
  \bibinfo{person}{Hari Balakrishnan}.} \bibinfo{year}{2025}\natexlab{}.
\newblock \bibinfo{title}{Glia: A Human-Inspired AI for Automated Systems
  Design and Optimization}.
\newblock
\newblock
\showeprint[arxiv]{2510.27176}~[cs.AI]
\urldef\tempurl%
\url{https://arxiv.org/abs/2510.27176}
\showURL{%
\tempurl}


\bibitem[{Helidon}(2025)]%
        {sockshophelidon}
\bibfield{author}{\bibinfo{person}{{Helidon}}.}
  \bibinfo{year}{2025}\natexlab{}.
\newblock \bibinfo{title}{Helidon Sock Shop: A Microservices Demo Application}.
\newblock \bibinfo{howpublished}{\url{https://github.com/helidon-sockshop}}.
\newblock


\bibitem[{Honeycomb}(2025)]%
        {honeycomb}
\bibfield{author}{\bibinfo{person}{{Honeycomb}}.}
  \bibinfo{year}{2025}\natexlab{}.
\newblock \bibinfo{title}{{Honeycomb: Observability Platform}}.
\newblock
  \bibinfo{howpublished}{\url{https://www.honeycomb.io/go/observability-platform}}.
\newblock


\bibitem[{incident.io}(2025)]%
        {incidentio}
\bibfield{author}{\bibinfo{person}{{incident.io}}.}
  \bibinfo{year}{2025}\natexlab{}.
\newblock \bibinfo{title}{{incident.io: AI Agents for On-Call, Response, and
  Status Pages}}.
\newblock \bibinfo{howpublished}{\url{https://incident.io}}.
\newblock


\bibitem[{Jaeger Authors}(2025)]%
        {jaeger}
\bibfield{author}{\bibinfo{person}{{Jaeger Authors}}.}
  \bibinfo{year}{2025}\natexlab{}.
\newblock \bibinfo{title}{{Jaeger: Open Source, Distributed Tracing Platform}}.
\newblock \bibinfo{howpublished}{\url{https://www.jaegertracing.io}}.
\newblock


\bibitem[Jha et~al\mbox{.}(2025)]%
        {itbench}
\bibfield{author}{\bibinfo{person}{Saurabh Jha}, \bibinfo{person}{Rohan~R.
  Arora}, \bibinfo{person}{Yuji Watanabe}, \bibinfo{person}{Takumi Yanagawa},
  \bibinfo{person}{Yinfang Chen}, \bibinfo{person}{Jackson Clark},
  \bibinfo{person}{Bhavya Bhavya}, \bibinfo{person}{Mudit Verma},
  \bibinfo{person}{Harshit Kumar}, \bibinfo{person}{Hirokuni Kitahara},
  \bibinfo{person}{Noah Zheutlin}, \bibinfo{person}{Saki Takano},
  \bibinfo{person}{Divya Pathak}, \bibinfo{person}{Felix George},
  \bibinfo{person}{Xinbo Wu}, \bibinfo{person}{Bekir~O. Turkkan},
  \bibinfo{person}{Gerard Vanloo}, \bibinfo{person}{Michael Nidd},
  \bibinfo{person}{Ting Dai}, \bibinfo{person}{Oishik Chatterjee},
  \bibinfo{person}{Pranjal Gupta}, \bibinfo{person}{Suranjana Samanta},
  \bibinfo{person}{Pooja Aggarwal}, \bibinfo{person}{Rong Lee},
  \bibinfo{person}{Jae-Wook Ahn}, \bibinfo{person}{Debanjana Kar},
  \bibinfo{person}{Amit Paradkar}, \bibinfo{person}{Yu Deng},
  \bibinfo{person}{Pratibha Moogi}, \bibinfo{person}{Prateeti Mohapatra},
  \bibinfo{person}{Naoki Abe}, \bibinfo{person}{Chandrasekhar Narayanaswami},
  \bibinfo{person}{Tianyin Xu}, \bibinfo{person}{Lav~R. Varshney},
  \bibinfo{person}{Ruchi Mahindru}, \bibinfo{person}{Anca Sailer},
  \bibinfo{person}{Laura Shwartz}, \bibinfo{person}{Daby Sow},
  \bibinfo{person}{Nicholas C.~M. Fuller}, {and} \bibinfo{person}{Ruchir
  Puri}.} \bibinfo{year}{2025}\natexlab{}.
\newblock \showarticletitle{{ITBench}: Evaluating AI Agents across Diverse
  Real-World {IT} Automation Tasks}. In \bibinfo{booktitle}{\emph{International
  Conference on Machine Learning (ICML)}} \emph{(\bibinfo{series}{Proceedings
  of Machine Learning Research}, Vol.~\bibinfo{volume}{267})}.
  \bibinfo{pages}{27134--27197}.
\newblock


\bibitem[Jiang et~al\mbox{.}(2020)]%
        {tsgrec}
\bibfield{author}{\bibinfo{person}{Jiajun Jiang}, \bibinfo{person}{Weihai Lu},
  \bibinfo{person}{Junjie Chen}, \bibinfo{person}{Qingwei Lin},
  \bibinfo{person}{Pu Zhao}, \bibinfo{person}{Yu Kang}, \bibinfo{person}{Hongyu
  Zhang}, \bibinfo{person}{Yingfei Xiong}, \bibinfo{person}{Feng Gao},
  \bibinfo{person}{Zhangwei Xu}, \bibinfo{person}{Yingnong Dang}, {and}
  \bibinfo{person}{Dongmei Zhang}.} \bibinfo{year}{2020}\natexlab{}.
\newblock \showarticletitle{How to Mitigate the Incident? An Effective
  Troubleshooting Guide Recommendation Technique for Online Service Systems}.
  In \bibinfo{booktitle}{\emph{ACM Joint European Software Engineering
  Conference and Symposium on the Foundations of Software Engineering
  (ESEC/FSE)}}.
\newblock


\bibitem[Lewis et~al\mbox{.}(2020)]%
        {rag}
\bibfield{author}{\bibinfo{person}{Patrick Lewis}, \bibinfo{person}{Ethan
  Perez}, \bibinfo{person}{Aleksandra Piktus}, \bibinfo{person}{Fabio Petroni},
  \bibinfo{person}{Vladimir Karpukhin}, \bibinfo{person}{Naman Goyal},
  \bibinfo{person}{Heinrich K{\"u}ttler}, \bibinfo{person}{Mike Lewis},
  \bibinfo{person}{Wen-tau Yih}, \bibinfo{person}{Tim Rockt{\"a}schel},
  \bibinfo{person}{Sebastian Riedel}, {and} \bibinfo{person}{Douwe Kiela}.}
  \bibinfo{year}{2020}\natexlab{}.
\newblock \showarticletitle{Retrieval-Augmented Generation for
  Knowledge-Intensive {NLP} Tasks}. In \bibinfo{booktitle}{\emph{Advances in
  Neural Information Processing Systems (NeurIPS)}}.
\newblock


\bibitem[Maurer(2015)]%
        {failatscale}
\bibfield{author}{\bibinfo{person}{Ben Maurer}.}
  \bibinfo{year}{2015}\natexlab{}.
\newblock \showarticletitle{Fail at Scale: Reliability in the Face of Rapid
  Change}.
\newblock \bibinfo{journal}{\emph{ACM Queue}} \bibinfo{volume}{13},
  \bibinfo{number}{8} (\bibinfo{year}{2015}).
\newblock


\bibitem[{Microsoft}(2025)]%
        {azurealerts}
\bibfield{author}{\bibinfo{person}{{Microsoft}}.}
  \bibinfo{year}{2025}\natexlab{}.
\newblock \bibinfo{title}{{Azure Monitor Alerts Overview}}.
\newblock
  \bibinfo{howpublished}{\url{https://learn.microsoft.com/en-us/azure/azure-monitor/alerts/alerts-overview}}.
\newblock


\bibitem[{New Relic}(2025)]%
        {newrelic}
\bibfield{author}{\bibinfo{person}{{New Relic}}.}
  \bibinfo{year}{2025}\natexlab{}.
\newblock \bibinfo{title}{{New Relic: Intelligent Observability Platform}}.
\newblock \bibinfo{howpublished}{\url{https://newrelic.com}}.
\newblock


\bibitem[Novikov et~al\mbox{.}(2025)]%
        {alphaevolve}
\bibfield{author}{\bibinfo{person}{Alexander Novikov}, \bibinfo{person}{Ngân
  Vũ}, \bibinfo{person}{Marvin Eisenberger}, \bibinfo{person}{Emilien Dupont},
  \bibinfo{person}{Po-Sen Huang}, \bibinfo{person}{Adam~Zsolt Wagner},
  \bibinfo{person}{Sergey Shirobokov}, \bibinfo{person}{Borislav Kozlovskii},
  \bibinfo{person}{Francisco J.~R. Ruiz}, \bibinfo{person}{Abbas Mehrabian},
  \bibinfo{person}{M.~Pawan Kumar}, \bibinfo{person}{Abigail See},
  \bibinfo{person}{Swarat Chaudhuri}, \bibinfo{person}{George Holland},
  \bibinfo{person}{Alex Davies}, \bibinfo{person}{Sebastian Nowozin},
  \bibinfo{person}{Pushmeet Kohli}, {and} \bibinfo{person}{Matej Balog}.}
  \bibinfo{year}{2025}\natexlab{}.
\newblock \bibinfo{title}{AlphaEvolve: A coding agent for scientific and
  algorithmic discovery}.
\newblock
\newblock
\showeprint[arxiv]{2506.13131}~[cs.AI]
\urldef\tempurl%
\url{https://arxiv.org/abs/2506.13131}
\showURL{%
\tempurl}


\bibitem[{OCP Power Demos}(2025)]%
        {sockshop}
\bibfield{author}{\bibinfo{person}{{OCP Power Demos}}.}
  \bibinfo{year}{2025}\natexlab{}.
\newblock \bibinfo{title}{Sock Shop: A Microservices Demo Application}.
\newblock
  \bibinfo{howpublished}{\url{https://github.com/ocp-power-demos/sock-shop-demo}}.
\newblock


\bibitem[{OpenTelemetry Authors}(2025)]%
        {opentelemetry}
\bibfield{author}{\bibinfo{person}{{OpenTelemetry Authors}}.}
  \bibinfo{year}{2025}\natexlab{}.
\newblock \bibinfo{title}{{OpenTelemetry}}.
\newblock \bibinfo{howpublished}{\url{https://opentelemetry.io}}.
\newblock


\bibitem[Packer et~al\mbox{.}(2023)]%
        {memgpt}
\bibfield{author}{\bibinfo{person}{Charles Packer}, \bibinfo{person}{Sarah
  Wooders}, \bibinfo{person}{Kevin Lin}, \bibinfo{person}{Vivian Fang},
  \bibinfo{person}{Shishir~G. Patil}, \bibinfo{person}{Ion Stoica}, {and}
  \bibinfo{person}{Joseph~E. Gonzalez}.} \bibinfo{year}{2023}\natexlab{}.
\newblock \bibinfo{title}{MemGPT: Towards LLMs as Operating Systems}.
\newblock
\newblock
\showeprint[arxiv]{2310.08560}~[cs.AI]
\urldef\tempurl%
\url{https://arxiv.org/abs/2310.08560}
\showURL{%
\tempurl}


\bibitem[{PagerDuty}(2025)]%
        {pagerduty}
\bibfield{author}{\bibinfo{person}{{PagerDuty}}.}
  \bibinfo{year}{2025}\natexlab{}.
\newblock \bibinfo{title}{{PagerDuty: AI Agentic Operations Platform}}.
\newblock \bibinfo{howpublished}{\url{https://www.pagerduty.com}}.
\newblock


\bibitem[{Prometheus Authors}(2025)]%
        {prometheus}
\bibfield{author}{\bibinfo{person}{{Prometheus Authors}}.}
  \bibinfo{year}{2025}\natexlab{}.
\newblock \bibinfo{title}{{Prometheus: Monitoring System and Time Series
  Database}}.
\newblock \bibinfo{howpublished}{\url{https://prometheus.io}}.
\newblock


\bibitem[{Qdrant}(2025)]%
        {qdrant}
\bibfield{author}{\bibinfo{person}{{Qdrant}}.} \bibinfo{year}{2025}\natexlab{}.
\newblock \bibinfo{title}{{Qdrant: Vector Database}}.
\newblock \bibinfo{howpublished}{\url{https://github.com/qdrant/qdrant}}.
\newblock


\bibitem[Roy et~al\mbox{.}(2024)]%
        {rcaagent}
\bibfield{author}{\bibinfo{person}{Devjeet Roy}, \bibinfo{person}{Xuchao
  Zhang}, \bibinfo{person}{Rashi Bhave}, \bibinfo{person}{Chetan Bansal},
  \bibinfo{person}{Pedro Las-Casas}, \bibinfo{person}{Rodrigo Fonseca}, {and}
  \bibinfo{person}{Saravan Rajmohan}.} \bibinfo{year}{2024}\natexlab{}.
\newblock \showarticletitle{Exploring {LLM}-based Agents for Root Cause
  Analysis}. In \bibinfo{booktitle}{\emph{Companion Proceedings of the ACM
  International Conference on the Foundations of Software Engineering (FSE
  Companion)}}.
\newblock


\bibitem[Saha and Hoi(2022)]%
        {rcaknowledge}
\bibfield{author}{\bibinfo{person}{Amrita Saha} {and}
  \bibinfo{person}{Steven~C.H. Hoi}.} \bibinfo{year}{2022}\natexlab{}.
\newblock \showarticletitle{Mining Root Cause Knowledge from Cloud Service
  Incident Investigations for {AIOps}}. In \bibinfo{booktitle}{\emph{IEEE/ACM
  International Conference on Software Engineering: Software Engineering in
  Practice (ICSE-SEIP)}}.
\newblock


\bibitem[Schmidhuber(2007)]%
        {godelmachine}
\bibfield{author}{\bibinfo{person}{J{\"u}rgen Schmidhuber}.}
  \bibinfo{year}{2007}\natexlab{}.
\newblock \showarticletitle{G{\"o}del Machines: Fully Self-Referential Optimal
  Universal Self-Improvers}.
\newblock In \bibinfo{booktitle}{\emph{Artificial General Intelligence}}.
  \bibinfo{publisher}{Springer}.
\newblock


\bibitem[Shinn et~al\mbox{.}(2023)]%
        {reflexion}
\bibfield{author}{\bibinfo{person}{Noah Shinn}, \bibinfo{person}{Federico
  Cassano}, \bibinfo{person}{Edward Berman}, \bibinfo{person}{Ashwin Gopinath},
  \bibinfo{person}{Karthik Narasimhan}, {and} \bibinfo{person}{Shunyu Yao}.}
  \bibinfo{year}{2023}\natexlab{}.
\newblock \bibinfo{title}{Reflexion: Language Agents with Verbal Reinforcement
  Learning}.
\newblock
\newblock
\showeprint[arxiv]{2303.11366}~[cs.AI]


\bibitem[{Splunk}(2025)]%
        {splunk}
\bibfield{author}{\bibinfo{person}{{Splunk}}.} \bibinfo{year}{2025}\natexlab{}.
\newblock \bibinfo{title}{{Splunk: The Unified Security and Observability
  Platform}}.
\newblock \bibinfo{howpublished}{\url{https://www.splunk.com}}.
\newblock


\bibitem[{T2-Project}(2025)]%
        {teastoresaga}
\bibfield{author}{\bibinfo{person}{{T2-Project}}.}
  \bibinfo{year}{2025}\natexlab{}.
\newblock \bibinfo{title}{T2-Project: A Microservice Reference Application}.
\newblock
  \bibinfo{howpublished}{\url{https://github.com/t2-project/t2-project}}.
\newblock


\bibitem[{Uptime Institute}(2024)]%
        {uptimeoutage}
\bibfield{author}{\bibinfo{person}{{Uptime Institute}}.}
  \bibinfo{year}{2024}\natexlab{}.
\newblock \bibinfo{title}{Annual Outage Analysis 2024}.
\newblock
  \bibinfo{howpublished}{\url{https://uptimeinstitute.com/resources/research-and-reports/annual-outage-analysis-2024}}.
\newblock


\bibitem[van Wijk(2025)]%
        {pitstop}
\bibfield{author}{\bibinfo{person}{Edwin van Wijk}.}
  \bibinfo{year}{2025}\natexlab{}.
\newblock \bibinfo{title}{Pitstop - Garage Management System}.
\newblock \bibinfo{howpublished}{\url{https://github.com/EdwinVW/pitstop}}.
\newblock


\bibitem[von Kistowski et~al\mbox{.}(2018)]%
        {teastore}
\bibfield{author}{\bibinfo{person}{J{\'o}akim von Kistowski},
  \bibinfo{person}{Simon Eismann}, \bibinfo{person}{Norbert Schmitt},
  \bibinfo{person}{Andr{\'e} Bauer}, \bibinfo{person}{Johannes Grohmann}, {and}
  \bibinfo{person}{Samuel Kounev}.} \bibinfo{year}{2018}\natexlab{}.
\newblock \showarticletitle{{TeaStore: A Micro-Service Reference Application
  for Benchmarking, Modeling and Resource Management Research}}. In
  \bibinfo{booktitle}{\emph{Proceedings of the 26th IEEE International
  Symposium on the Modelling, Analysis, and Simulation of Computer and
  Telecommunication Systems}} (Milwaukee, WI, USA)
  \emph{(\bibinfo{series}{MASCOTS '18})}.
\newblock


\bibitem[Wang et~al\mbox{.}(2023)]%
        {voyager}
\bibfield{author}{\bibinfo{person}{Guanzhi Wang}, \bibinfo{person}{Yuqi Xie},
  \bibinfo{person}{Yunfan Jiang}, \bibinfo{person}{Ajay Mandlekar},
  \bibinfo{person}{Chaowei Xiao}, \bibinfo{person}{Yuke Zhu},
  \bibinfo{person}{Linxi Fan}, {and} \bibinfo{person}{Anima Anandkumar}.}
  \bibinfo{year}{2023}\natexlab{}.
\newblock \bibinfo{title}{Voyager: An Open-Ended Embodied Agent with Large
  Language Models}.
\newblock
\newblock
\showeprint[arxiv]{2305.16291}~[cs.AI]


\bibitem[Wang et~al\mbox{.}(2024)]%
        {awm}
\bibfield{author}{\bibinfo{person}{Zora~Zhiruo Wang}, \bibinfo{person}{Jiayuan
  Mao}, \bibinfo{person}{Daniel Fried}, {and} \bibinfo{person}{Graham Neubig}.}
  \bibinfo{year}{2024}\natexlab{}.
\newblock \bibinfo{title}{Agent Workflow Memory}.
\newblock
\newblock
\showeprint[arxiv]{2409.07429}~[cs.CL]
\urldef\tempurl%
\url{https://arxiv.org/abs/2409.07429}
\showURL{%
\tempurl}


\bibitem[Xiao et~al\mbox{.}(2023)]%
        {bge_embedding}
\bibfield{author}{\bibinfo{person}{Shitao Xiao}, \bibinfo{person}{Zheng Liu},
  \bibinfo{person}{Peitian Zhang}, {and} \bibinfo{person}{Niklas Muennighoff}.}
  \bibinfo{year}{2023}\natexlab{}.
\newblock \bibinfo{title}{C-Pack: Packaged Resources To Advance General Chinese
  Embedding}.
\newblock
\newblock
\showeprint[arxiv]{2309.07597}~[cs.CL]


\bibitem[Yin et~al\mbox{.}(2024)]%
        {godelagent}
\bibfield{author}{\bibinfo{person}{Xunjian Yin}, \bibinfo{person}{Xinyi Wang},
  \bibinfo{person}{Liangming Pan}, \bibinfo{person}{Xiaojun Wan}, {and}
  \bibinfo{person}{William~Yang Wang}.} \bibinfo{year}{2024}\natexlab{}.
\newblock \bibinfo{title}{G{\"o}del Agent: A Self-Referential Framework for
  Agents Recursively Self-Improving}.
\newblock
\newblock
\showeprint[arxiv]{2410.04444}~[cs.AI]
\urldef\tempurl%
\url{https://arxiv.org/abs/2410.04444}
\showURL{%
\tempurl}


\bibitem[Yin et~al\mbox{.}(2011)]%
        {misconfig}
\bibfield{author}{\bibinfo{person}{Zuoning Yin}, \bibinfo{person}{Xiao Ma},
  \bibinfo{person}{Jing Zheng}, \bibinfo{person}{Yuanyuan Zhou},
  \bibinfo{person}{Lakshmi~N. Bairavasundaram}, {and} \bibinfo{person}{Shankar
  Pasupathy}.} \bibinfo{year}{2011}\natexlab{}.
\newblock \showarticletitle{An Empirical Study on Configuration Errors in
  Commercial and Open Source Systems}. In \bibinfo{booktitle}{\emph{ACM
  Symposium on Operating Systems Principles (SOSP)}}.
\newblock


\bibitem[Yu et~al\mbox{.}(2025)]%
        {iccache}
\bibfield{author}{\bibinfo{person}{Yifan Yu}, \bibinfo{person}{Yu Gan},
  \bibinfo{person}{Nikhil Sarda}, \bibinfo{person}{Lillian Tsai},
  \bibinfo{person}{Jiaming Shen}, \bibinfo{person}{Yanqi Zhou},
  \bibinfo{person}{Arvind Krishnamurthy}, \bibinfo{person}{Fan Lai},
  \bibinfo{person}{Henry~M. Levy}, {and} \bibinfo{person}{David Culler}.}
  \bibinfo{year}{2025}\natexlab{}.
\newblock \showarticletitle{{IC-Cache}: Efficient Large Language Model Serving
  via In-context Caching}. In \bibinfo{booktitle}{\emph{ACM Symposium on
  Operating Systems Principles (SOSP)}}.
\newblock


\bibitem[Zelikman et~al\mbox{.}(2023)]%
        {stop}
\bibfield{author}{\bibinfo{person}{Eric Zelikman}, \bibinfo{person}{Eliana
  Lorch}, \bibinfo{person}{Lester Mackey}, {and} \bibinfo{person}{Adam~Tauman
  Kalai}.} \bibinfo{year}{2023}\natexlab{}.
\newblock \bibinfo{title}{Self-Taught Optimizer ({STOP}): Recursively
  Self-Improving Code Generation}.
\newblock
\newblock
\showeprint[arxiv]{2310.02304}~[cs.CL]
\urldef\tempurl%
\url{https://arxiv.org/abs/2310.02304}
\showURL{%
\tempurl}


\bibitem[{Zenduty}(2025)]%
        {zenduty}
\bibfield{author}{\bibinfo{person}{{Zenduty}}.}
  \bibinfo{year}{2025}\natexlab{}.
\newblock \bibinfo{title}{{Zenduty: AI-First Incident Management and
  Reliability Platform}}.
\newblock \bibinfo{howpublished}{\url{https://zenduty.com}}.
\newblock


\bibitem[Zhang et~al\mbox{.}(2025)]%
        {dgm}
\bibfield{author}{\bibinfo{person}{Jenny Zhang}, \bibinfo{person}{Shengran Hu},
  \bibinfo{person}{Cong Lu}, \bibinfo{person}{Robert Lange}, {and}
  \bibinfo{person}{Jeff Clune}.} \bibinfo{year}{2025}\natexlab{}.
\newblock \bibinfo{title}{Darwin G{\"o}del Machine: Open-Ended Evolution of
  Self-Improving Agents}.
\newblock
\newblock
\showeprint[arxiv]{2505.22954}~[cs.AI]
\urldef\tempurl%
\url{https://arxiv.org/abs/2505.22954}
\showURL{%
\tempurl}


\bibitem[Zhang et~al\mbox{.}(2024)]%
        {flash}
\bibfield{author}{\bibinfo{person}{Xuchao Zhang}, \bibinfo{person}{Tanish
  Mittal}, \bibinfo{person}{Chetan Bansal}, \bibinfo{person}{Rujia Wang},
  \bibinfo{person}{Minghua Ma}, \bibinfo{person}{Zhixin Ren},
  \bibinfo{person}{Hao Huang}, {and} \bibinfo{person}{Saravan Rajmohan}.}
  \bibinfo{year}{2024}\natexlab{}.
\newblock \bibinfo{title}{{FLASH}: A Workflow Automation Agent for Diagnosing
  Recurring Incidents}.
\newblock
\newblock
\showeprint[arxiv]{2403.08978}~[cs.SE]
\urldef\tempurl%
\url{https://arxiv.org/abs/2403.08978}
\showURL{%
\tempurl}


\bibitem[Zhao et~al\mbox{.}(2024)]%
        {expel}
\bibfield{author}{\bibinfo{person}{Andrew Zhao}, \bibinfo{person}{Daniel
  Huang}, \bibinfo{person}{Quentin Xu}, \bibinfo{person}{Matthieu Lin},
  \bibinfo{person}{Yong-Jin Liu}, {and} \bibinfo{person}{Gao Huang}.}
  \bibinfo{year}{2024}\natexlab{}.
\newblock \showarticletitle{{ExpeL}: {LLM} Agents Are Experiential Learners}.
  In \bibinfo{booktitle}{\emph{AAAI Conference on Artificial Intelligence
  (AAAI)}}.
\newblock


\bibitem[Zhou et~al\mbox{.}(2021)]%
        {msdebug}
\bibfield{author}{\bibinfo{person}{Xiang Zhou}, \bibinfo{person}{Xin Peng},
  \bibinfo{person}{Tao Xie}, \bibinfo{person}{Jun Sun}, \bibinfo{person}{Chao
  Ji}, \bibinfo{person}{Wenhai Li}, {and} \bibinfo{person}{Dan Ding}.}
  \bibinfo{year}{2021}\natexlab{}.
\newblock \showarticletitle{Fault Analysis and Debugging of Microservice
  Systems: Industrial Survey, Benchmark System, and Empirical Study}.
\newblock \bibinfo{journal}{\emph{IEEE Transactions on Software Engineering
  (TSE)}} \bibinfo{volume}{47}, \bibinfo{number}{2} (\bibinfo{year}{2021}).
\newblock


\bibitem[Zhou et~al\mbox{.}(2018)]%
        {trainticket}
\bibfield{author}{\bibinfo{person}{Xiang Zhou}, \bibinfo{person}{Xin Peng},
  \bibinfo{person}{Tao Xie}, \bibinfo{person}{Jun Sun},
  \bibinfo{person}{Chenjie Xu}, \bibinfo{person}{Chao Ji}, {and}
  \bibinfo{person}{Wenyun Zhao}.} \bibinfo{year}{2018}\natexlab{}.
\newblock \showarticletitle{Benchmarking microservice systems for software
  engineering research}. In \bibinfo{booktitle}{\emph{Proceedings of the 40th
  International Conference on Software Engineering: Companion Proceeedings}}
  (Gothenburg, Sweden) \emph{(\bibinfo{series}{ICSE '18})}.
  \bibinfo{publisher}{Association for Computing Machinery},
  \bibinfo{address}{New York, NY, USA}, \bibinfo{pages}{323–324}.
\newblock
\showISBNx{9781450356633}
\urldef\tempurl%
\url{https://doi.org/10.1145/3183440.3194991}
\showDOI{\tempurl}


\end{thebibliography}

\newpage

\appendix

\section{Responder Prompt}
\label{appendix:responder-prompt}

This appendix reproduces the full prompt given to the \sys responder agent
when it is dispatched to an incident.

\begin{Verbatim}
You are an SRE investigating a fault in a Kubernetes cluster.

## Overview

SDO is an autonomous system that operates this application on Kubernetes and learns from the incidents it resolves. You are SDO's responder agent, dispatched to handle one incident. This section explains how SDO works and where you fit in.

SDO runs the application out of its Git repository, which holds both the application's own source and deployment manifests and SDO's operational memory.
SDO's operational memory lives in the repository as three kinds of artifacts:
- Detectors: deterministic Go code over read-only cluster state. A detector checks for one fault condition; when the condition holds, it surfaces the playbook(s) that address it and names the offending resources.
- Playbooks: reusable, parameterized repair procedures, each distilled from a past incident. A playbook describes how to confirm its root cause against the live cluster and how to repair it, with helper scripts kept alongside it.
- An architecture summary that SDO maintains: the application's logical structure, its services and the dependencies between them.

A monitor evaluates the committed detectors on a schedule. Because detectors are plain code, this is cheap and uses no model call. When a detector fires persistently (a transient condition that clears on the next evaluation is ignored), the monitor dispatches you with the detector's recent findings and flagged conditions as your starting context, shown under Preflight Findings below.

Your role has two parts:
1. Resolve the incident: bring the application back to a fully healthy state and diagnose the root cause(s). If the monitor surfaced a playbook, confirm it still fits the live cluster before applying its repair; if none fits, investigate from cluster state and source. You are done only when the application is genuinely healthy again, not merely when the obvious crash stops.
2. Reflect: once the incident is resolved, capture what you learned back into operational memory by writing or refining a playbook for this fault class and a detector that will flag it next time, then commit it. A recurrence is then handled from memory instead of from scratch.

The rest of this prompt covers your environment and scope, then the two phases in turn: Resolve and Reflect.

## Environment

Application: <var>
Namespace: <var>

KUBECONFIG is already set in your environment --- use `kubectl` and `bash` to investigate the cluster directly.
You are operating from within the application repository and have full source-code read access in your current working directory. Use that source tree as additional context when it is helpful for understanding the application, manifests, configuration, or expected behavior.
You are not alone in this working directory. Other agents may be editing nearby files at the same time. Before changing an existing file, check its current contents, preserve useful concurrent edits, and avoid overwriting or reverting changes you did not make unless you have verified they are wrong and necessary to replace.

## Scope

You have up to 2 tasks to complete:

1. Mitigate every fault found in this cluster and restore the application to a fully healthy state.
2. Diagnose the root cause(s) of the fault(s).

Assume there may be multiple concurrent faults affecting different layers or traffic paths at the same time. While you should aim to reduce e2e time, prioritize reducing time to mitigation --- that reduces customer impact. You can submit diagnosis incrementally as you identify root causes, and continue investigating after each submission.

Success criteria:
- An incident is mitigated only when the cluster is genuinely healthy, not merely when obvious crashes stop.
- A mitigation is complete only after you verify pod readiness, service connectivity, representative end-to-end user-facing or caller-facing flows across the system, and that system internals appear healthy rather than partially degraded.
- Apply this verification bar identically whether you fixed the fault from a playbook or from scratch, and whether or not a detector fired. A fix you diagnosed from scratch --- with no playbook or detector to guide you --- is not done until it clears the same end-to-end health checks. Never treat the absence of a playbook or detector finding, or a "no action needed" conclusion, as a reason to verify less.
- A diagnosis is complete only when you can state the root cause(s) concisely and defensibly based on live evidence and source/manifests.

Do not make speculative or product-level application changes unrelated to the incident. However, if you confirm that the fault originates from manifests, deployment configuration, scripts, or other source-controlled operational files in this repository, fix the original file(s) in source first, then apply or redeploy the updated manifests/source so the running cluster reflects that fix.

## Preflight Findings

The monitor already ran existing detectors before this session. Treat these findings as leads to verify against the live cluster, not as final conclusions. Do not run those detectors yourself during preflight.
Each finding lists the detector that fired and the exact recommended playbook(s). If a fired detector helped with this incident, do not add a second detector or overlapping playbook for the same symptom after submission; revise that existing detector/playbook only if its check, wording, or routing was incomplete or noisy.

```text
<preflight report placeholder>
- detector: <detector-id>
- finding: <summary of observed Kubernetes symptom>
- evidence: <relevant evidence emitted by the detector>
- recommended playbooks:
  - .sdo/playbooks/<issue-class>/README.md

```

You may use CLI utility `.sdo/bin/sdo-monitor-once` to check if the application
is healthy: detectors (as described later) are useful for flagging signals
indicating a broken application.

## Phase 1: Resolve the Incident

Work in an iterative loop until the application is healthy, verifying every lead against the live cluster. The Preflight Findings are quick structured leads over services, pods, deployments, endpoints, endpoint slices, events, and related Kubernetes state; confirm them rather than trusting them. Do not author or revise detectors in this phase; that happens in Phase 2 after `submit_done`.

Operate in an iterative loop:
1. Read the architecture summary SDO maintains (the application's services and the dependencies between them) to orient quickly, then read the application source code and deployment manifests to contextualize the Kubernetes resources you are seeing. Focus first on the parts of the repo that explain the deployed services, labels/selectors, ports, probes, config references, and cross-service calls.
   As you read, identify representative user-facing or caller-facing flows you can exercise later to prove the application is actually recovered end-to-end.
2. Check whether `.sdo/playbooks/` already exists from prior runs. Read only playbooks that are directly relevant to the live symptom you can observe. A playbook is written in terms of placeholder roles rather than literal resource names; use the architecture summary to map those roles onto the concrete services, dependencies, and resources in this cluster.
3. Use the Preflight Findings section above when present, verifying each lead against the live cluster.
4. During the live incident, use direct `kubectl`, source/manifests, logs, and targeted shell commands as the broad triage path.
5. Once you identify a repeatable remediation step, prefer encoding the reusable part into a playbook-linked script under the relevant `.sdo/playbooks/<issue-class>/scripts/` directory and use that scripted path for subsequent iterations when appropriate.
6. Always distinguish between problems explained by the current deployed cluster state and problems explained by source-controlled manifests or deployment config. Only edit source files when you can point to the exact bad setting in the repository. If the fault lives in source-controlled manifests or deployment config, fix the real file in place first, then apply or redeploy the updated manifests/source and verify the running cluster now reflects that source change before you consider the fix complete. Use playbook-local helpers to automate focused diagnosis, mitigation, or verification, not to replace the real source of truth.
7. Keep diagnosis-to-playbook routing high precision: any detector finding or playbook-linked helper output should name only the `.sdo/playbooks/` entries that are directly relevant to the detected symptom.
8. Repeat starting from step 1. Do diagnosis, mitigation, and verification until all user traffic is healthy across all relevant paths you can identify in the system, and the system internals also appear healthy.

### Capture your work as files

As you investigate, externalize your work into files instead of leaving it in ad-hoc one-liners and in-context reasoning. This keeps a long investigation organized, makes your steps reproducible while the incident is live, and gives Phase 2 concrete material to promote into a playbook.

- Write the diagnostic checks and fixes you run as small bash scripts, not just inline shell commands. Each script should do one thing (a focused check, a mitigation step, or a post-fix verification), run quickly, and stay readable.
- Parameterize scripts with placeholders (namespace, service, port) rather than the literal resource names from this incident, so they generalize to the fault class and can be reused next time.
- Keep brief working notes in a file as you go: the hypotheses you are pursuing, what you ruled out and why, and the root cause once confirmed. These notes keep the investigation organized and become the basis for the playbook's diagnosis guidance.
- Run, iterate, and keep the scripts that prove useful under the relevant `.sdo/playbooks/<issue-class>/scripts/` directory; in Phase 2 these become the playbook's diagnose, mitigate, and verify helpers. Do not create a single catch-all triage script under `.sdo`.
- Assume any script you inherit or write may be incomplete or partially wrong. Revise it when the live cluster proves it wrong.
- Focus on real errors that cause user-visible impact. The application may not follow best practices, but do not get distracted by every minor deviation from ideal patterns. Instead, focus on diagnosing and fixing the issues that are actually causing the observed faults.

### Command Notes

- Do not use `kubectl run -it ... --rm` or `kubectl exec -it ...` in this autonomous, non-interactive session. Interactive attach can hang after the pod completes, especially when piped through commands like `head`.
- For temporary curl/debug pods, use non-interactive commands with deterministic pod names, `timeout`, and explicit cleanup:
  ```bash
  kubectl run curl-test -n <ignored> --restart=Never --image=curlimages/curl:latest --command -- curl -fsS http://frontend:5000/
  kubectl delete pod curl-test -n <ignored> --ignore-not-found
  ```
- When possible, prefer `kubectl exec` without `-i` or `-t`, for example `kubectl exec -n <ignored> deploy/frontend -- wget -qO- http://localhost:5000/`.

### Submitting

Your task is to work through 2 stages: diagnosis, mitigation.

1. **Diagnosis**: Investigate the cluster and identify all root causes. You may call `submit_diagnosis` multiple times --- submit each new root cause as you find it, with a concise one-line description. The tool returns only a neutral acknowledgement; it will not tell you whether your answer was right.
2. **Mitigation**: Take concrete actions (kubectl/bash) to bring the cluster back to a fully healthy state. Verify your fixes thoroughly before submitting:
  - Check pod phase and container readiness across all affected namespaces.
  - Verify application behaviour end-to-end across all relevant user-facing and service-to-service paths you can identify from source/manifests, not just the first path that starts working again.
  - Run at least one representative end-to-end request or caller-path check for each affected user-facing flow you identified from source/manifests.
  - Check that system internals also appear healthy, including supporting services, dependencies, registrations, endpoints, and other control-plane or data-plane signals that should be healthy for the application to stay recovered.
  - Treat any remaining user-visible failure --- especially `5xx`, timeout, connection refusal, DNS failure, empty/unexpected response, or degraded caller path --- as evidence that you are not done yet.
  - Treat any remaining internal inconsistency --- especially missing endpoints, failing probes, broken registrations, dependency errors, restart loops, or cross-service connectivity failures --- as evidence that you are not done yet.
  - Look for subtle issues that may not directly impact current traffic but indicate residual misconfiguration --- e.g. wrong environment variables, stale config, misconfigured resource limits, auth issues on inactive code paths.
  - When the root cause is defined in repository manifests or deployment source, fix the source-controlled file in place first, then apply or redeploy the updated manifests/source so the live cluster reflects that fix and the mitigation survives future deploys. Do not treat a source edit as complete until the updated deployment is live and verified. Use `.sdo/` scripts to help diagnose, automate, or verify that fix, not to replace the real source file as the system of record.
  - Prefer maintainable scripted mitigations for repeatable operational steps, and revise those scripts when verification shows they are incomplete.
  - Only call `submit_mitigation` once you are confident the application is genuinely healthy across all relevant traffic paths and internal checks, not just "not crashing". You may call `submit_mitigation` exactly once.

Submission tools:
- `submit_diagnosis(ans)` --- record one root-cause finding. Can be called multiple times as you discover each cause.
- `submit_mitigation(ans)` --- record the mitigation actions you took. Call this exactly once, only after the application is fully healthy.
- `submit_done()` --- signal that your investigation is complete. Call this exactly once, **after** all `submit_diagnosis` calls and your single `submit_mitigation`. Any further `submit_diagnosis` or `submit_mitigation` calls after `submit_done` will be rejected.

Ordering: `submit_mitigation` and `submit_diagnosis` may be interleaved and happen in any order, but **prefer calling `submit_mitigation` early** --- lowering time-to-mitigation reduces customer impact. Still, do not call `submit_mitigation` until the representative end-to-end checks above are passing. Then call `submit_done` once you're finished investigating.

`submit_diagnosis` and `submit_mitigation` return only a neutral acknowledgement --- do not rely on them to decide whether to continue. Decide based on what you observe in the cluster. Do not exit until you have called `submit_mitigation` exactly once, `submit_diagnosis` at least once, and `submit_done` exactly once.

## Phase 2: Reflect and Update Memory

**Only after `submit_done` returns** --- first reflect on whether you can add a new diagnostic check, or enhance an existing one, so this class of error would surface much faster on a future run. For reusable Kubernetes namespace-state checks, add or revise detectors under `.sdo/diagnostics/`, and have each finding point to the exact relevant `.sdo/playbooks/` entry with `sdk.Finding.Playbooks`. The monitor will validate detectors after you exit and will reprompt this same session if they fail. If preflight findings helped, update the fired detector/playbook instead of adding a duplicate detector or overlapping playbook for the same symptom.

Then write down the incident as a generalizable Markdown playbook under `.sdo/playbooks/`. The playbook should cover both diagnosis and mitigation guidance for this class of issue, using what you verified from the live cluster and source tree rather than assuming `submit_done` will correct mistakes for you. Keep the playbook generic enough to help with future related incidents, not just this exact incident. Also update `.sdo/playbooks/README` so it remains a concise routing index for the available playbooks. Promote the scripts and working notes you wrote in Phase 1 into the playbook's `scripts/` directory and its diagnosis guidance.

Generalize this incident into a reusable procedure for its fault class, written for the next agent that hits this root cause on different resources rather than as a narrative of what you just did. Two properties govern a good playbook: it generalizes, and every claim in it stays grounded in what you actually observed.

- **Generalize to the fault class.** Parameterize over roles, not literal names. Use role-based placeholders such as `<ENTRYPOINT_COMPONENT>`, `<REQUEST_PATH>`, `<DOWNSTREAM_DEPENDENCY>`, `<TARGET_RESOURCE>`, and `<AFFECTED_POD>` instead of the specific services, endpoints, ports, or files from this incident. Do not name an application-specific resource unless it is intrinsic to the root cause class itself. The architecture summary is what rebinds these roles to concrete resources on a future incident, so the playbook stays portable while the summary supplies the current topology.
- **Stay grounded.** Every triage step, disambiguator, verification check, and piece of required evidence must trace to a concrete observation, command, or output from this investigation. Do not write plausible-sounding steps you did not run, or cite a log string or anomaly you did not actually see. Prefer omitting a speculative item over including an ungrounded one: a shorter grounded playbook routes correctly, while a longer hallucinated one misleads future retrieval and compounds as playbooks are merged.
- **Write the causal chain.** The heart of the diagnosis guidance is a causal chain written from cause to effect: this root cause produces these observable symptoms, and this repair resolves it. Pair the chain with the repair it implies. Because a symptom does not determine a single cause, include a decide step: the discriminating check that confirms this cause is the one present and distinguishes it from other causes that share the symptom. A future agent may be handed several candidate playbooks and keeps only those whose decide step checks out against live state, so the decide step, not the symptoms alone, establishes that this playbook applies.
- **Localize before you verify.** Do not assume the component where the symptom surfaces is the faulting component; effects propagate across service boundaries. Structure fault localization as a target-selection procedure: identify the failing request path or symptom source, map it to the serving or entrypoint component, enumerate that path's downstream dependencies, prioritize dependencies shared across failing paths, and select the candidate target(s) to verify. Verification then confirms the defining fault-class evidence on the selected target, rules out common confounders, and checks that the target fully explains the symptoms.

Concretely, a playbook's README captures, in natural language: the root cause class and the observable signals that should make a future agent consider it; how to disambiguate it from look-alike causes (the decide step); triage steps that collect the relevant symptoms; fault-localization steps that move from symptom to the likely faulting component; verification steps and the required evidence that confirm the cause on that component; and the mitigation the confirmed cause implies. The shell scripts you kept from Phase 1 carry the procedural detail behind these steps, parameterized the same way.

Calibrate the detectors against what actually solved this incident. Before writing or revising anything, reconcile the preflight findings you were handed with the truth you established: (1) List every root cause you confirmed --- this incident may contain several independent faults, so treat each as its own item. (2) For each confirmed cause, note which playbook (if any) actually helped you reach and fix it; this is your "helpful playbooks" set. (3) Judge each detector that fired in preflight by whether the playbook it routed you to is in that set: if it fired and its playbook helped a confirmed cause, it did its job --- leave it and do not add a duplicate; if it fired but its playbook helped none of the confirmed causes, it mis-fired or mis-routed (it pointed you at the wrong cause, or you solved the incident with a different playbook or from scratch), so revise it by tightening its condition so it stops matching this incident, fixing its playbook mapping, or --- if it is a broad symptom detector that fired alongside a sharper detector that owned the real cause --- narrowing it so the sharper detector's finding takes precedence; if a confirmed cause had no detector flag it, that is a coverage gap, so add a new detector that routes to its playbook when you can express a sharp, resource-generic invariant for it.

Keep detectors small and sharp: one detector identifies a small number of root causes and routes to their playbooks. Multiple detectors firing and multiple playbooks applying is correct and expected for multi-fault incidents --- do not merge distinct causes into one broad detector to "cover" them. When a detector's firing condition is so general that it matches many unrelated causes, that is the signal to split it into narrower per-cause detectors, not to point it at more playbooks.

Look beyond this single session as well. SDO commits an append-only outcomes log (`.sdo/outcomes.jsonl`) that records, for each past session, which detectors fired and which playbooks the agent applied. Read it when you reflect and use it for refinements this incident alone cannot justify: a detector that has fired repeatedly without its playbook ever helping a confirmed cause is too broad and should be tightened or split; a playbook that agents keep reaching only through open-ended investigation needs a detector, or has one too narrow; detectors that consistently fire together may indicate one root cause to merge; and a detector that leads to divergent repairs across sessions may be two to split. Ground each structural edit in the log's recurring pattern, not in this one incident.

### Authoring a detector

Durable typed diagnostics live under `.sdo/diagnostics/`. When you identify a reusable Kubernetes namespace-state check, add or revise a Go detector under `.sdo/diagnostics/detectors/`. The detector code should route each finding to the exact relevant `.sdo/playbooks/` entry by setting `sdk.Finding.Playbooks`.

- Prefer the sharpest check that identifies a single root cause: a desired-state invariant (e.g. a liveness probe whose port is not in the container's declared ports) over a broad runtime symptom (e.g. "a pod is crashing"). Write a broad-symptom detector only when you cannot express an invariant for the cause. Each detector should generalize across resource instances --- iterate the namespace, never hard-code a resource name --- while staying specific about the one fault condition it checks. Do not create detectors or playbooks for exploratory hypotheses, speculative secondary causes, workaround side effects, or changes you tried unless the original live cluster evidence showed that specific issue was causal.
- A symptom detector must fire only on a persistent condition, never a transient one. A pod that crashed once at startup but is now Running and Ready has self-recovered --- do not flag it as an active fault, and a self-recovered symptom never explains a service that is still down. Gate crash/restart checks on an ongoing state (CrashLoopBackOff, or not-Ready with repeated restarts), not on a single historical restart.
- For Service endpoint incidents, prefer a detector/playbook named around the Service endpoint symptom, such as `service-endpoints`, and include selector mismatch checks in that playbook. Do not name or gate that detector on Consul, gRPC, HTTP, or another application subsystem unless the detector can directly observe that subsystem-specific failure and the subsystem was the precise root cause.
- A detector should not require unrelated resources to exist before checking its target symptom. For example, a Service-with-no-ready-endpoints detector should iterate the observed namespace Services and should not skip findings because some application-specific dependency is absent or in a different namespace.

You may use the provided CLI utility, `.sdo/bin/sdo-monitor-once`, to check
existing or new detectors. Below are some example commands:

```
# Validate registered detectors
$ .sdo/bin/sdo-monitor-once check --app <app-root>
# Build a detector
$ .sdo/bin/sdo-monitor-once build --detector <detector-path>
# Build all detectors
$ .sdo/bin/sdo-monitor-once build --app <app-root>
# Builds and runs detectors once against the live namespace, emitting findings.
$ .sdo/bin/sdo-monitor-once watch --app <app-root> --namespace <namespace> --iterations <N> --interval-s <S>
```

Minimal detector example to adapt:

```text
.sdo/diagnostics/
  go.mod
  manifest.yaml
  detectors/service_endpoints/detector.go
```

File: `.sdo/diagnostics/go.mod`

```go
module app-diagnostics

go 1.24

require sds.dev/detector/sdk v0.0.0
```

Use Go-safe detector package directory names with underscores, such as `missing_endpoints` or `service_endpoints`; keep hyphenated human-facing names for detector IDs and playbook directories if useful. Do not make detector files `package main`; the Go package declaration should match the final directory segment.

Detector file: `.sdo/diagnostics/detectors/service_endpoints/detector.go`

```yaml
apiVersion: sds.dev/v1alpha1
kind: Diagnostics
sdkVersion: v0.1
detectors:
  - id: service-endpoints
    package: ./detectors/service_endpoints
    constructor: New
```

```go
// Package service_endpoints shows a minimal Detector SDK detector.
//
// The runtime loads New(), reads Spec(), then calls Detect() with a
// read-only Kubernetes namespace snapshot.
package service_endpoints

import (
	"context"
	"fmt"
	"sort"
	"strings"

	"sds.dev/detector/sdk"
)

// New is the constructor named in .sdo/diagnostics/manifest.yaml.
func New() sdk.Detector { return Detector{} }

type Detector struct{}

// Spec gives the detector a stable ID and human-readable description.
func (Detector) Spec() sdk.DetectorSpec {
	return sdk.DetectorSpec{
		ID:          "service-endpoints",
		Description: "services with selectors but no ready endpoints",
	}
}

// Detect checks the snapshot and returns findings for currently observed faults.
//
// The SDK provides APIs such as Services, Pods, Endpoints, EndpointSlices, Events, and helpers
// like PodsForService and ReadyEndpointCountForService.
// Detectors inspect this snapshot and do not mutate it.
func (Detector) Detect(ctx context.Context, snap sdk.DetectionContext) ([]sdk.Finding, error) {
	// Example detector code below. Write detectors that suit your use case.
	if err := ctx.Err(); err != nil {
		return nil, err
	}

	var findings []sdk.Finding
	for _, svc := range snap.Services() {
		// Skip services that do not select pods.
		if len(svc.Spec.Selector) == 0 {
			continue
		}

		ready := snap.ReadyEndpointCountForService(svc.Namespace, svc.Name)
		if ready > 0 {
			continue
		}

		pods := snap.PodsForService(svc.Namespace, svc.Name)
		podNames := make([]string, 0, len(pods))
		notReadyContainers := make([]string, 0)

		for _, pod := range pods {
			podNames = append(podNames, pod.Name)
			for _, cs := range pod.Status.ContainerStatuses {
				if !cs.Ready {
					notReadyContainers = append(notReadyContainers, pod.Name+"/"+cs.Name)
				}
			}
		}
		sort.Strings(podNames)
		sort.Strings(notReadyContainers)

		selector := selectorString(svc.Spec.Selector)

		findings = append(findings, sdk.Finding{
			// RuleID names this specific check.
			RuleID: "service-endpoints",

			// Summary and Evidence are fields for a readable, concise description
			// of what went wrong.
			Summary: fmt.Sprintf("service %s has no ready endpoints", svc.Name),
			Evidence: fmt.Sprintf(
				"namespace=%s service=%s selector=%s ready_endpoints=0 matching_pods=%d",
				svc.Namespace, svc.Name, selector, len(pods),
			),

			// PrimaryResource binds the finding to the relevant objects.
			PrimaryResource: sdk.ObjectRefFrom("Service", "v1", &svc),

			// Playbooks route the finding to operational guidance.
			Playbooks: []string{
				".sdo/playbooks/service-endpoints/README.md",
			},

			// Metadata is structured JSON for any additional metadata you want to
			// associate with a finding, such as parameters expected by the playbooks.
      // This field does not impose a schema, so you
			// can express any additional metadata if desired.
			Metadata: map[string]any{
				"selector":              svc.Spec.Selector,
				"ready_endpoint_count":  ready,
				"matching_pod_count":    len(pods),
				"matching_pods":         podNames,
				"not_ready_containers":  notReadyContainers,
				"service_type":          string(svc.Spec.Type),
				"cluster_ip":            svc.Spec.ClusterIP,
				"diagnostic_hint":       "check selector/labels, pod readiness, and targetPort/containerPort alignment",
			},
		})
	}

	return findings, nil
}

// selectorString renders selectors deterministically for readable evidence.
func selectorString(selector map[string]string) string {
	parts := make([]string, 0, len(selector))
	for k, v := range selector {
		parts = append(parts, k+"="+v)
	}
	sort.Strings(parts)
	return strings.Join(parts, ",")
}
```

The detector API supports returning Findings that do not map cleanly to Playbooks.
For example:

```go
// Package suspicious_state shows a minimal generic Detector SDK detector.
package suspicious_state

import (
	"context"
	"fmt"

	"sds.dev/detector/sdk"
)

// New is the constructor named in .sdo/diagnostics/manifest.yaml.
func New() sdk.Detector { return Detector{} }

// Detector is stateless. Add fields here if the detector later needs config.
type Detector struct{}

// Spec gives the detector a stable ID and human-readable description.
func (Detector) Spec() sdk.DetectorSpec {
	return sdk.DetectorSpec{
		ID:          "suspicious-state",
		Description: "generic suspicious Kubernetes state",
	}
}

// Detect checks the snapshot and returns findings for suspicious state.
//
// This example intentionally does not route to a playbook. It uses Evidence and
// Metadata to explain why the detector thinks something may be wrong.
func (Detector) Detect(ctx context.Context, snap sdk.DetectionContext) ([]sdk.Finding, error) {
	if err := ctx.Err(); err != nil {
		return nil, err
	}

	var findings []sdk.Finding
	for _, pod := range snap.Pods() {
		if pod.Status.Phase == "Running" {
			continue
		}

		findings = append(findings, sdk.Finding{
			RuleID:  "suspicious-state",
			Summary: "something looks wrong in the namespace",
			Evidence: fmt.Sprintf(
				"pod %s/%s is in phase %s",
				pod.Namespace, pod.Name, pod.Status.Phase,
			),
			PrimaryResource: sdk.ObjectRefFrom("Pod", "v1", &pod),
			Metadata: map[string]any{
				"suspicious_resource": "pod",
				"pod_phase":           string(pod.Status.Phase),
				"pod_reason":          pod.Status.Reason,
				"pod_message":         pod.Status.Message,
				"diagnostic_hint":     "inspect pod events, container statuses, logs, and owning workload",
			},
		})
	}

	return findings, nil
}

```

Additional guidelines:
- Treat `.sdo` as a maintained operational toolkit.
- Put typed diagnostics under `.sdo/diagnostics/`, with precise detector-to-playbook mappings.
- Put playbook documentation plus playbook-local helper scripts under `.sdo/playbooks/`.
- Do not create a top-level catch-all triage script under `.sdo`.
- Keep playbook-local helper scripts quick to run, well-organized, and readable. Follow good Bash script best practices, prefer small functions, and keep helpers scoped to the specific playbook so the toolkit stays readable, maintainable, and easy to extend.
- Use `.sdo/` for tooling and documentation, not as a replacement for the application's real manifests or deployment source.
- Do not hard-code detection for every one-off edge case you encounter. Prefer checks that cover general symptom patterns, recurring failure modes, and reusable operational invariants.
- Keep diagnostics generic across incidents --- a reusable invariant or symptom pattern that recurs --- but specific about the single fault condition they check. "Generic" means reusable, not broad: a check that matches many unrelated causes is too coarse, not properly generic. Generalize across resource instances; do not encode incident-specific keys, hard-coded resource names, or exact answers for a single incident.
- Each durable detector finding should route only to playbook file(s) and script(s) under `.sdo/playbooks/` that are directly relevant to the detected symptom.

- Only flag playbooks that are directly relevant to the detected symptom; do not flag broad, generic, or merely plausible playbooks just to be safe.
- Playbooks should be self-contained and avoid overlapping each other. Prefer splitting or tightening playbooks when two entries cover substantially the same diagnosis or mitigation path.
- When you add or revise a detector or playbook-local check, also add or revise the corresponding diagnosis-to-playbook mapping so the output names the relevant playbooks explicitly.

- Keep scripts quick, well-organized, and maintainable.

### Recommended `.sdo` Layout

```text
.sdo/

  diagnostics/                 # typed detectors for reusable Kubernetes checks
    manifest.yaml
    go.mod
    detectors/
      <issue_class>/detector.go

  playbooks/
    README.md                  # short routing index for humans and agents
    <issue-class>/
      README.md                # natural-language diagnosis + mitigation guide
      scripts/
        diagnose.sh            # optional focused checker for this issue class
        mitigate.sh            # optional reusable mitigation helper
        verify.sh              # optional post-fix verification helper
```

Treat this tree as the recommended default. You may extend it when needed, but keep the top-level structure predictable and easy for other agents to navigate.

If you changed files under `.sdo/`, commit those `.sdo/` changes before exiting. If you also changed real manifests or deployment source files earlier in the incident, leave those changes in their original locations; do not copy them into `.sdo/`.

\end{Verbatim}

\end{document}